\documentclass[twocolumn,superscriptaddress,amssymb,amsmath,nobibnotes,aps,prd,showpacs,nofootinbib]{revtex4-1}
\pdfoutput=1
\usepackage{epsfig}
\usepackage{xcolor}
\usepackage{graphicx,subfigure,bm,color,psfrag,hyperref}
\usepackage{amsfonts}
\usepackage{lipsum}
\usepackage{caption}
\usepackage{mathtools}
\usepackage{ulem}

\newcommand{\be}{\begin{equation}}
\newcommand{\ee}{\end{equation}}

\usepackage{tabularx}

\makeatletter
\renewcommand{\p@subsection}{}
\renewcommand{\p@subsubsection}{}
\makeatother

\begin{document}

\title{Anisotropic massive Brans-Dicke gravity extension of the standard $\Lambda$CDM model}

\author{\"{O}zg\"{u}r Akarsu}
\email{akarsuo@itu.edu.tr}
\affiliation{Department of Physics, \. Istanbul Technical University, Maslak 34469 \. Istanbul, Turkey}

\author{Nihan Kat{\i}rc{\i}}
\email{nihan.katirci@itu.edu.tr}
\affiliation{Department of Physics, \. Istanbul Technical University, Maslak 34469 \. Istanbul, Turkey}

\author{Ne\c se \" Ozdemir}
\email{nozdemir@itu.edu.tr}
\affiliation{Department of Physics, \. Istanbul Technical University, Maslak 34469 \. Istanbul, Turkey}

\author{J. Alberto V\'azquez}
\email{javazquez@icf.unam.mx}
\affiliation{Instituto de Ciencias F\'isicas, Universidad Nacional Aut\'onoma de M\'exico, Apdo. Postal 48-3, 62251 Cuernavaca, Morelos, M\'exico}

\begin{abstract}
We present an explicit detailed theoretical and observational investigation of an anisotropic massive Brans-Dicke (BD) gravity extension of the standard $\Lambda$CDM model, wherein the extension is characterized by two additional degrees of freedom; the BD parameter, $\omega$, and the present day density parameter corresponding to the shear scalar, $\Omega_{\sigma^2,0}$. The BD parameter, determining the deviation from general relativity (GR), by alone characterizes both the dynamics of the effective dark energy (DE) and the redshift dependence of the shear scalar. These two affect each other depending on $\omega$, namely, the shear scalar contributes to the dynamics of the effective DE, and its anisotropic stress --which does not exist in scalar field models of DE within GR-- controls the dynamics of the shear scalar deviating from the usual $\propto(1+z)^6$ form in GR. We mainly confine the current work to non-negative $\omega$ values as it is the right sign --theoretically and observationally-- for investigating the model as a correction to the $\Lambda$CDM. By considering the current cosmological observations, we find that $\omega\gtrsim 250$, $\Omega_{\sigma^2,0}\lesssim 10^{-23}$ and the contribution of the anisotropy of the effective DE to this value is insignificant. We conclude that the simplest anisotropic massive BD gravity extension of the standard $\Lambda$CDM model exhibits no significant deviations from it all the way to the Big Bang Nucleosynthesis. We also point out the interesting features of the model in the case of negative $\omega$ values; for instance, the constraints on $\Omega_{\sigma^2,0}$ could be relaxed considerably, the values of $\omega\sim-1$ (relevant to string theories) predict dramatically different dynamics for the expansion anisotropy.
\end{abstract}

\maketitle
\section{Introduction}
\label{sec:intro}
 The base $\Lambda$-cold dark matter ($\Lambda$CDM) model, relying on the inflationary paradigm \cite{Starobinsky:1980te,Starobinsky:1980teER,Guth:1980zm,Guth:1980zmER,Linde:1981mu,Linde:1981muER,Albrecht:1982wi,Albrecht:1982wiER}, is the simplest and most successful cosmological model to describe the dynamics and the large scale structure in agreement with the most of the currently available observational data \cite{Ade:2015xua,Aghanim:2018eyx,Alam:2016hwk,Abbott:2017wau}. It is today credited as the \textit{standard cosmological model}, yet it is probably not where the story has concluded but the hardest part has just begun. It suffers from severe theoretical issues relating to the cosmological constant $\Lambda$ being responsible for the late time acceleration of the Universe \cite{Weinberg:1988cp,Sahni:1999gb,Peebles:2002gy,Padmanabhan:2002ji,Velten:2014nra,Padilla:2015aaa,Lopez-Corredoira:2017rqn} and, on the observational side, from tensions of various degrees of significance between some existing data sets \cite{tension01,tension02, tension03,Delubac:2014aqe,Aubourg:2014yra,Riess:2016jrr,Zhao:2017cud,Bullock:2017xww,Freedman:2017yms}. Besides, based on the most minimal a priori assumptions, model independent reconstructions of the evolution of the dark energy (DE) equation of state (EoS) parameter $w$ \cite{Vazquez:2012ce,Hee:2016ce,Wang:2018fng} and also model independent diagnoses \cite{Sahni:2014ooa,Shafieloo:2018gin} exhibit a dynamical behaviour of $w(z)$.

 Nevertheless, even small deviations from/corrections to the $\Lambda$CDM mostly imply/require profound, and sometimes highly non-trivial, modifications to the fundamental theories of physics \cite{Caldwell:2009ix}. Indeed, we still do not have a promising and concrete fundamental theory leading to DE models that are more general than $\Lambda$ and also can account for the small, but significant, deviations from the $\Lambda$CDM model as persistently suggested by the high precision data. Though, depending on the characteristics of the DE models favored by observations, we can decide whether it is more natural to consider DE as an \textit{actual} physical source \cite{Peebles:2002gy,Copeland:2006wr} or as an \textit{effective} source originated from a modification \cite{DeFelice:2010aj,Clifton:2011jh,Capozziello:2011et,Bamba:2012cp,Nojiri:2017ncd} to the standard theory of gravity, i.e., general relativity (GR). For instance, eliminating $\Lambda$ by detecting $w >-1$ would not be by itself illuminating to the nature of DE, but any detection of $w < -1$ would be very illuminating to the nature of gravitation. For a perfect barotropic fluid, the adiabatic sound speed $c^2_a$ is the physical propagation speed of perturbations, and therefore $w<-1$ (viz., phantom \cite{Caldwell:1999ew} or quintom \cite{Guo:2004fq,Cai:2009zp} DE models) typically accompanying by $c^2_a<0$,  implies the instability of perturbations (and/or ghost instabilities, see, e.g., \cite{Dubovsky:2005xd}), whereas, as shown in \cite{Boisseau:2000pr,Gannouji:2006jm}, scalar-tensor theories of gravity, such as Brans-Dicke (BD) theory, can lead to an effective source with $w<-1$ that does not correspond to the change of the sign of $c^2_a$, and hence perturbations can still be stable. Similarly, any detection of DE with negative energy density would also be hugely informative, such that this for an actual physical source is of course physically ill, whereas it can be an effective source so that negative energy density does not lead to any pathology since this is not the true energy density. On the observational side, the constraints on the EoS parameter of DE persistently indicate that $w\approx -1$, and so do not exclude $w < -1$ \cite{Aghanim:2018eyx} and it has recently been shown in \cite{Aubourg:2014yra,Tamayo:2019,Sahni:2014ooa} that DE models with energy densities passing below zero at high redshifts fit the data better and can address the tensions relevant to Lyman-$\alpha$ forest measurements \cite{Delubac:2014aqe}. Such DE sources are indeed possible, in general, in modified theories of gravity with an effective gravitational coupling strength weaker in the past, e.g., in scalar-tensor theories, if we collect all modifications to the usual Einstein field equations to define an effective DE. Consequently, it can be argued that any detection of a deviation from $\Lambda$CDM model implies that the late time acceleration is not driven by $\Lambda$ \footnote{The converse is not true, even if it turns out that DE exactly yields $w = -1$.} and of DE yielding $w< -1$ and/or $\rho<0$ implies that gravity is not minimally coupled (which eliminate the actual perfect-fluid models of DE and stands as a strong sign in favor of the effective DE models from modified gravity).

Generically, the effective DE models from modified gravity induce non-zero anisotropic stresses\footnote{See \cite{Barrow:1997sy} for a list of well know anisotropic stresses and their effects on the expansion anisotropy.}, which can be realized if we relax the isotropic space assumption of the $\Lambda$CDM model along with the modification to GR. Of course, we are not able to observe anisotropic stresses directly, yet they can reveal themselves through their effect on the evolution of the expansion anisotropy as well as on the average expansion rate of the Universe. Then, a natural question is whether the observations allow or suggest an anisotropic space, unless otherwise anisotropic stresses would be irrelevant. Indeed, there are some clues for questioning the spatially maximally symmetric Universe assumption, i.e., Robertson-Walker (RW) background, of the $\Lambda$CDM model. This has been mainly motivated by hints of anomalies in the CMB distribution first observed on the full sky by the WMAP experiment \cite{deOliveira-Costa:2003utu,Eriksen:2003db,Vielva:2003et,Cruz:2004ce} and so remained in Planck experiment \cite{Ade:2013nlj,Ade:2013ydc,Ade:2013xla,Ade:2013vbw}. So far, the local deviations from the statistically highly isotropic Gaussianity of the CMB in some directions (the so called cold spots) could not have been excluded at high confidence levels \cite{Cruz:2004ce,Bennett11,Pontzen10,Ade:2013nlj}. Furthermore, it has been shown that the CMB angular power spectrum has a quadrupole power lower than expected from the best-fit $\Lambda$CDM model \cite{Efstathiou:2003tv,Ade:2013zuv}. Several explanations for this anomaly have been proposed \cite{DeDeo:2003te,Cline:2003ve,Tsujikawa:2003gh,Koivisto:2005mm,Campanelli:2006vb,Gruppuso:2007ya,Campanelli:2007qn,Rodrigues:2007ny,campanelli2011,Vazquez:2012} including the anisotropic expansion of the Universe. On the other hand, inflation (canonical) isotropizes the Universe very efficiently \cite{Wald:1983ky,Starobinsky:1982mr}, leaving a residual anisotropy that is negligible for any practical application in the observable Universe. This could be irrelevant if an anisotropic expansion is developed only well after the matter-radiation decoupling, for example during the domination of DE, say, by means of its anisotropic pressure acting as a late-time source of not insignificant anisotropy \cite{Chimento:2005ua,Battye:2006mb,Koivisto:2007bp,Rodrigues:2007ny,Koivisto:2008ig,Koivisto:2008xf,Akarsu:2008fj,Cooray:2008qn,Campanelli:2009tk,Akarsu:2013dva,Koivisto:2014gia,Koivisto:2015vda} (see also, e.g., \cite{Mota:2007sz,appleby10,Appleby:2012as,Amendola:2013qna} for constraint studies on the anisotropy of DE). Indeed, the CMB provides very tight constraints on the anisotropy at the time of recombination \cite{Martinez95,Bunn:1996ut,Kogut:1997az} of the order of the quadrupole temperature, i.e., $(\Delta T/T)_{\ell=2} \sim 10^{-5}$. And, in the simplest anisotropic generalization of the standard $\Lambda$CDM (viz., replacing the spatially flat RW metric by Bianchi type I metric), the energy density corresponding to the expansion anisotropy scales as the inverse of the square of the comoving volume, $\rho_{\sigma^2}\propto S^{-6}$, which implies an isotropization of the expansion from the recombination up to the present, leading to the typically derived upper bounds on the corresponding density parameter today as $\Omega_{\sigma^2,0}\sim 10^{-20}$. However, this is true if the anisotropic expansion is not generated by any anisotropic source, say, by an anisotropic DE (effective or actual), arising after decoupling \cite{Chimento:2005ua,Battye:2006mb,Koivisto:2007bp,Rodrigues:2007ny,Koivisto:2008ig,Koivisto:2008xf,Akarsu:2008fj,Cooray:2008qn,Campanelli:2009tk,Akarsu:2013dva,Koivisto:2014gia,Koivisto:2015vda}. Nevertheless, almost all of these strong constraints on the expansion anisotropy in the late Universe, $\Omega_{\sigma^2,0}\lesssim10^{-22}$, assume the non-existence of anisotropic sources (actual or effective) in the late universe, and are derived in fact from the contribution of the expansion anisotropy on the average expansion rate of the Universe \cite{Barrow:1976rda,Pontzen16,Saadeh:2016bmp,Saadeh:2016sak,Akarsu:2019pwn}. On the other hand, direct observational constraints, e.g., from SNIa data, on the expansion anisotropy of the later Universe are much weaker, namely, on the order of $\Omega_{\sigma^2,0}\lesssim 10^{-4}$ for $z\sim0$ \cite{Campanelli:2010zx,Wang:2017ezt}.

It is in fact always possible to imitate any modification to the Einstein field equations as GR with an arbitrary mixture of scalars, vectors and tensors \cite{Hu:2007pj,Kunz08}. Therefore, one may think of abandoning any attempt to distinguish between the actual physical DE and the effective DE from modified gravities. However, a modified theory of gravity, e.g., the Brans-Dicke theory, can modify both the average expansion rate of the Universe and the evolution of the expansion anisotropy in a \textit{distinctive} way depending on its free parameter characterizing the model, e.g., the Brans-Dicke parameter $\omega$. So, of course, owing to Occam's razor, looking for these two concurrent modifications predicted by a specific modified theory of gravity and testing them against the observational data can provide us a strong reason for favoring or disfavoring the modified theory of gravity under consideration over GR. In this paper, for instance, the effective EoS parameter corresponding to the expansion anisotropy, the present day EoS of the effective DE and the redshift at which it evolves into the phantom region range in specific intervals as $1< w_{\sigma^2}\leq\frac{5}{3}$, $-1.14\lesssim w<-1$ and $0.50 \leq z_{\rm PDL}< 0.65$ (see  \cite{Boisseau:2010pd} for a similar result) for positive values of the BD parameter $\omega$.
Indeed, revealing the origin of the late time acceleration of the Universe considering such distinctive features of the DE models from modified theories of gravity is among the scientific themes underlying some upcoming experiments (e.g., \cite{Amendola:2016saw}).

Finally, we have also independent motivations for modifying GR by fundamental theoretical physics. Namely, almost all of the attempts to quantize GR introduce deviations from it in the form of extra degrees of freedom, higher powers of the curvature in the action, higher order derivatives in the field equations, or non-local terms. The low-energy limits of the string theories typically yield BD gravity with $\omega=-1$ and similarly $d$-brane models yield BD gravity with $\omega\sim-1$ \cite{Duff:1994an,Lidsey:1999mc,Callan:1985ia,Fradkin:1985ys}. Indeed, among all possible alternatives to GR, it is found natural by many to first consider scalar-tensor theories \cite{Jordan:1959eg,Brans:1961sx,Bergmann68,Nordvedt70,wagoner70}, which only add a (usually massive) scalar degree of freedom, the BD-like scalar field, to the two massless, spin-2 polarizations (gravitons) contained in the metric tensor. In relevance with our focus in this paper, the field equations in such models can be regarded as effective Einstein field equations and so the terms originating from the BD-like scalar field and its derivatives, moved to the right hand side, can always be interpreted as an effective fluid. Within this approach, the correspondence between general scalar-tensor theories and this effective fluid has been worked out explicitly first in \cite{Pimentel89} (see also \cite{Madsen88} for the preceding work), and has shown that, in general, the corresponding effective fluid is imperfect. A recent work \cite{Faraoni:2018qdr} extended and completed the correspondence showing how a symmetry of BD theory translates into a symmetry of this fluid such that this correspondence is valid for any spacetime geometries. Brans and Dicke's 1961 theory \cite{Jordan:1959eg,Brans:1961sx}, motivated first by the implementation of Mach's principle in gravity, today provides a prototype of scalar-tensor theories \cite{Bergmann68,Nordvedt70,wagoner70}.

The original BD theory  \cite{Jordan:1959eg,Brans:1961sx} has only one additional constant parameter $\omega$, determining the deviations from GR, which is recovered in the limit of $|\omega| \rightarrow \infty$. Theoretically $\omega\geq-\frac{3}{2}$ is necessary to avoid the Jordan/scalar field ($\varphi$) from yielding negative energy density values in the Einstein frame that leads, e.g., the Minkowski vacuum to be unstable. Observations on a wide range of scales further constrain BD theory around GR. The tightest constraints, to date, are imposed by the observations of gravitational phenomena in the Solar System; specifically, $\omega \gtrsim 40000$ at the 2$\sigma$ confidence level (C.L.) from observations of radio signals from the Cassini spacecraft as it passes behind the Sun \cite{Bertotti:2003rm}. It can be vastly relaxed for the massive BD theory, i.e., when the Jordan field is accompanied by a potential $U(\varphi)$, as in this case one should consider the corresponding mass ($M$) of the Jordan field along with the BD parameter, viz., the $\{\omega,M\}$ parameter region \cite{Perivolaropoulos:2009ak}. It is shown that $\omega=O(1)$ is allowed for the Solar System constraints provided that $M\gtrsim 2\times 10^{-17}\rm eV$ \cite{Perivolaropoulos:2009ak}. The constraints on $M$, however, are typically subject to the cosmological observations, since, in particular, it becomes cosmologically significant at substantially lower scales, viz., at the Hubble mass scale $M_{H_{0}}\simeq 10^{-33}\rm eV$, which leaves the constraint from the Cassini spacecraft still valid.

The cosmological constraints on $\omega$ are complementary to the local constraints as they probe different length and time scales, as well as different epochs of the Universe. The strongest cosmological constraint, which assumes $U(\varphi)=\rm const.$ and considers CMB data from the first Planck release along with the constraints from Big Bang Nucleosynthesis (BBN) light element abundances, gives $\omega>890$ at the 2$\sigma$ C.L. \cite{Avilez:2013dxa}. The constraints from Planck 2015 (2013) and a compilation of baryon acoustic oscillations (BAO) data are $\omega>333$ ($\omega>208$) at the 2$\sigma$ C.L., weakly dependent on the exponent $n>0$ for a power-law potential $U(\varphi)\propto \varphi^n$ mimicking $\Lambda$ in the late universe \cite{Ballardini:2016cvy,Umilta:2015cta} (also see \cite{Rossi:2019lgt} for a recent work considering a more complicated potential). The BBN leads to $\omega\gtrsim 277$ (for a universe containing dust, radiation and conventional vacuum energy), which assumes power-law solutions and, by using general solutions, can be somewhat altered depending on the behaviour of the Jordan field in the early universe \cite{Clifton:2005xr}. It is forecast that future cosmological experiments will be able to constrain $\omega$ at a level comparable to the local tests \cite{Acquaviva:2007mm,Alonso:2016suf,Ballardini:2019tho}. 

 The typical mass scale of $M\sim 10^{-33}\rm eV$ imposed by cosmological observations is so small that the local constraint $\omega \gtrsim 40000$ from the Cassini spacecraft must be satisfied at all times in all parts of the Universe, and leads to the conclusion that the massive BD extension of the standard $\Lambda$CDM model must be phenomenologically very similar to it throughout most of the history of the Universe, implying that the cosmological features of BD theory would be observationally irrelevant. On the other hand, it is still worth studying in detail and understanding the cosmological constructions within massive BD theory by considering the cosmological constraints on $\omega$ only, as BD theory is representative of some more general gravity theories that can evade the local constraints (see \cite{Clifton:2011jh} and references therein for further reading). For instance, straightforwardly, it is possible to consider a further extension of BD theory by allowing the BD parameter to be some functions of the Jordan field, $\omega=\omega(\varphi)$, in addition to the presence of potential $U(\varphi)$ and this leads to the possibility of having an $\omega$ small enough to be significant in the past, while being large enough today to be consistent with the tight constraints imposed by observations of gravitational phenomena in the Solar System. Alternatively, for instance, in the Horndeski theory (the most general scalar-tensor theory having second-order field equations in four dimensions) \cite{Horndeski:1974wa,Deffayet:2011gz}, while on cosmological scales BD theory emerges as an approximation, the derivative self-interactions of the Horndeski scalar can be large enough on small scales (higher curvature than cosmological environments) to screen the scalar resulting at the same time to the recovery GR. For example, it was proposed in \cite{Nicolis:2008in} that the so called Galileon theories (a close cousin of BD theory) could also explain the late time acceleration of the Universe while evading solar system constraints. A different kind of example is that, in the presence of extra dimensions, it is possible to make BD theory (massive or massless) consistent with gravitational tests (including solar system tests) for $|\omega|=O(1)$ \cite{Akarsu:2019oem}. See also \cite{Ishak:2018his} for a recent overview on cosmological probes of gravity for testing deviations from GR in relevance with such modified gravity theories.

Motivated by the discussions above, we present a detailed theoretical and observational investigation of the anisotropic (LRS Bianchi type I metric) massive Brans-Dicke gravity extension of the standard $\Lambda$CDM model, which is characterized by two additional free parameters, namely, the BD parameter $\omega$ and the corresponding present day density parameter of the expansion anisotropy $\Omega_{\sigma^2,0}$. The role of the cosmological constant is taken over by the Jordan field potential of the form $U(\varphi)\propto\varphi^2$ and the physical ingredient of the Universe is considered as in the standard $\Lambda$CDM. We confine the present work, basically, to the non-negative values of the BD parameter since, here, we mainly consider the \textit{extension} as a \textit{correction} to the standard $\Lambda$CDM.

\section{Anisotropic massive Brans-Dicke gravity extension of the $\Lambda$CDM model}
\label{sec:LCDMsol}
We consider the BD action \cite{Jordan:1959eg,Brans:1961sx} written in the Jordan frame in the following form:
\begin{equation}
\begin{aligned}
\label{eq:action}
S_{\rm JBD}=&\int {\rm d}^4 x \sqrt{-g}\bigg[\frac{\varphi^2}{8}R-\omega\bigg(\frac{1}{2}\nabla_{\mu}\varphi \nabla^{\mu}\varphi+\frac{1}{2}M^2\varphi^2\bigg)\bigg]\\
&+S_{\rm Matter},
\end{aligned}
\end{equation}
where $\varphi
=\varphi(t)$ is the Jordan scalar field (function of cosmic time $t$ only) and $\omega=\rm const.$ is the Brans-Dicke parameter, $R$ is the Ricci scalar, $g$ is the determinant of the metric $g_{\mu\nu}$, and $S_{\rm Matter}$ is the matter action, which is independent of $\varphi$ so that the weak equivalence principle is satisfied. It is clear from the way of writing the action that the term $M^2$ stands as the \textit{bare} mass-squared of the Jordan field.\footnote{We consider the bare mass as it is done in \cite{Boisseau:2010pd}, where it formally appears as the mass of a minimally coupled canonical scalar field when curvature scalar is dropped. On the other hand, when the \textit{effective mass} of the Jordan field is considered there are various definitions in the literature (see, e.g., \cite{Faraoni:2009km} for a recent discussion and further references), though one cannot say that these would not fail from a strict particle physics point of view.} We assume $M^2={\rm const.}$ so that, as can also be seen from the action, it stands like a cosmological constant as $2\omega M^2\equiv\Lambda$ and thereby can drive accelerated expansion. Hence, switching to massive BD from GR with a positive cosmological constant provides us with an opportunity to construct $\Lambda$CDM-type cosmologies, such that the mass of the Jordan field alone can play the role of positive cosmological constant like in the standard $\Lambda$CDM cosmology provided that $2\omega M^2\equiv\Lambda>0$, and the Jordan field $\varphi$ varying slowly enough on the top of this can account for small deviations from the standard $\Lambda$CDM model in a particular way, which in turn may lead to an improved fit to the observational data w.r.t. the standard $\Lambda$CDM model. In line with that, we intend to study the BD extension of the standard $\Lambda$CDM model as a \textit{correction}, and therefore we demand the term $2\omega M^2$ to be positive definite, which requires $\omega>0$ as long as we keep $M^2>0$ to avoid the Jordan field from having an imaginary mass.\footnote{If this is not the case then $2\omega M^2=\Lambda$ will contribute to the field equations like a negative cosmological constant, which may be compensated by a rapidly changing Jordan field. Within the standard BD gravity (i.e., massless BD gravity) in the presence of pressureless source, there exist cosmological solutions with accelerating expansion if $-2<\omega<-1$, and Jordan field is real (i.e., the effective cosmological gravitational coupling is positive definite) as well if $-\frac{3}{2}<\omega<-\frac{4}{3}$ leading to deceleration parameter in the range $0>q>-1$, and to $\varphi^2\propto (1+z)^2$ and $\varphi^2\propto (1+z)^3$ in the two boundaries of this range, correspondingly \cite{WeinbergBook}. Hence, obviously, if we consider $\omega<0$ and $M^2>0$ leading to $2\omega M^2<0$ (i.e., effectively negative cosmological constant), then it will be necessary to confine ourselves to the range $-\frac{3}{2}<\omega<-\frac{4}{3}$, and moreover the decelerating effect of the term $2\omega M^2<0$ would be compensated by bringing the value of the BD parameter closer to $-4/3$, which in turn, implies a faster Jordan field [viz., for $\omega\sim -4/3$ Jordan field changes as fast as the pressureless source, i.e., $\varphi^2 \sim (1+z)^3$] whereas we are looking for slowly changing Jordan field, say, $\varphi^2\sim {\rm const.}$ since we demand it to do only small modifications on the $\Lambda$CDM dynamics.} Hence, in this study, unless otherwise is mentioned, we carry out our investigations by assuming $\omega>0$, which is already stronger than the assumption $\omega\geq -3/2$ to avoid the Jordan field from yielding negative energy density values in the Einstein frame that leads, for instance the Minkowski vacuum to be unstable.

We consider the simplest anisotropic generalization of the spatially flat and homogeneous spacetime, i.e., locally rotationally symmetric (LRS) Bianchi type I metric, which can be written as follows;
\begin{equation}
\label{metric}
 {\rm d}s^2 = -{\rm d} t^2+S^2\left[e^{4\beta}{\rm d}x^2+ e^{-2\beta}({\rm d}y^2+{\rm d}z^2)\right],
\end{equation}
where $S=S(t)$ is the mean scale factor. Here, the exponent $\beta=\beta(t)$ satisfies the relation $\dot{\beta}^2=\frac{1}{6}\sigma^2$, where $\sigma^2=\sigma_{ij}\sigma^{ij}$ and $\sigma^{ij}$  are shear scalar and tensor, respectively. The scalar curvature for this metric \eqref{metric} may be written in terms of the mean scale factor and shear scalar as $R=-6\left(\frac{\ddot S}{S}+\frac{\dot{S}^2}{S^2}\right)-\sigma^2$. Throughout the paper a dot denotes derivative w.r.t. cosmic time $t$.

We consider all types of matter distribution (namely, the usual cosmological sources such as radiation, baryons, etc.) as isotropic perfect fluids, which are described by the following energy-momentum tensor (EMT)
\begin{equation}
\label{eq:EMT}
{T_{\mu}}^{\nu}={\textnormal{diag}}[-\rho,p,p,p],
\end{equation}
where $\rho$ and $p$ are the energy density and pressure, respectively.

The field equations for the action \eqref{eq:action} within the framework of the metric \eqref{metric} in the presence of \eqref{eq:EMT} read:
 \begin{align}
 3\frac{\dot{S}^2}{S^2}-\frac{1}{2}\sigma^2-2\omega \frac{\dot{\varphi}^2}{\varphi^2}+6\frac{\dot{S}}{S}\frac{\dot{\varphi}}{\varphi}-2\omega M^2=\frac{4}{\varphi^2}\rho \label{eq:rho},
  \end{align}
  \begin{equation}
  \begin{aligned}
 &2\frac{\ddot S}{S}+\frac{\dot{S}^2}{S^2}+\frac{1}{2}\sigma^2-\sqrt{\frac{2}{3}}\left[\dot{\sigma}+\left(3\frac{\dot{S}}{S}+2\frac{\dot{\varphi}}{\varphi} \right)\sigma\right]\\
&+(2 \omega+2) \frac{\dot{\varphi}^2}{\varphi^2}
 +2\frac{\ddot{\varphi}}{\varphi}+4\frac{\dot{S}}{S}\frac{\dot{\varphi}}{\varphi}-2 \omega M^2=-\frac{4}{\varphi^2}p,
 \label{eq:pres1}
 \end{aligned}
 \end{equation}
 \begin{equation}
  \begin{aligned}
  &2\frac{\ddot S}{S}+\frac{\dot{S}^2}{S^2}+\frac{1}{2}\sigma^2+\sqrt{\frac{1}{6}}\left[\dot{\sigma}+\left(3\frac{\dot{S}}{S}+2\frac{\dot{\varphi}}{\varphi} \right)\sigma\right]\\
 &+(2 \omega+2) \frac{\dot{\varphi}^2}{\varphi^2}+2\frac{\ddot{\varphi}}{\varphi}+4\frac{\dot{S}}{S}
  \frac{\dot{\varphi}}{\varphi}-2 \omega M^2=-\frac{4}{\varphi^2}p,
  \label{eq:pres2}
   \end{aligned}
 \end{equation}
 for which the latter two correspond to the pressure equations along the $x$-axis and the $y$- and $z$-axes, respectively. The Klein-Gordon equation for the Jordan field reads
 \begin{equation}
\frac{\ddot{\varphi}}{\varphi}+3\frac{\dot{S}}{S}\frac{\dot{\varphi}}{\varphi} -\frac{3}{2\omega}\left(\frac{\ddot S}{S}+\frac{\dot{S}^2}{S^2}\right)-\frac{1}{2\omega}\frac{\sigma^2}{2}+M^2=0.
\label{eq:phi}
 \end{equation}
We consider the standard cosmological fluids, namely, pressureless fluid ($p_{\rm m}= 0$) and radiation/relativistic fluids ($p_{\rm r} = \rho_{\rm r}/3$), so that, here, $\rho=\rho_{\rm m}+\rho_{\rm r}$ and $p=p_{\rm r}$. If the Jordan field is approximately constant then the first Friedmann equation \eqref{eq:rho} becomes $3\frac{\dot{S}^2}{S^2}\approx\frac{4}{\varphi^2}\rho+\frac{\sigma^2}{2}+2\omega M^2$ (as like in GR), where the effective cosmological gravitational coupling strength is then given by $G=\frac{\varphi_0^2}{\varphi^2}\, G_0$, where $G=\frac{1}{2\pi\varphi^2}$ and $G_0=\frac{1}{2\pi\varphi^2_0}$ with zero subscript denoting today's value (throughout the paper).\footnote{Note however that for bound systems in the quasi-static regime, e.g., our solar system, the effective Newton's constant is $G_{\rm N}=\frac{2\omega+3}{2\omega+4}\,G_0$ \cite{Brans:1961sx}, thus observers in a bound system which formed today would measure the same cosmological and local gravitational strength if $\varphi^2=\frac{2\omega+4}{2\omega+3}\varphi_0^2$.}

We consider the conventional representation of the true equation for scalar-tensor gravity interacting with matter in the Einsteinian form with the constant $G_0=G(t_0)$
\begin{equation}
\label{einsteinian}
R_{\mu\nu}-\frac{1}{2}R g_{\mu\nu}=8\pi G_0 (T_{\mu\nu,{\rm m}}+T_{\mu\nu,{\rm DE}}),
\end{equation}
$8\pi G_0=4/\varphi^2_0$ following Refs. \cite{Boisseau:2000pr,Gannouji:2006jm}, where it is discussed  that modified gravity theories can be recast in the standard GR form such that \eqref{einsteinian} where all the new geometrical terms are grouped (on the r.h.s.) to form an effective DE contribution denoted as $T_{\mu\nu,{\rm DE}}$ (for a detailed discussion, see introduction in Ref. \cite{Gannouji:2006jm}). We note that the shear scalar is kept on the left hand side as it is a part of the Einstein tensor, namely, the metric itself only. Accordingly, we define the effective DE in the following way:
\begin{equation}
\begin{aligned}
\label{stdfried}
3\frac{\dot S^2}{S^2}-\frac{1}{2}\sigma^2=\frac{4}{\varphi^2_0}\,(\rho+\rho_{\rm DE}),
\end{aligned}
\end{equation}
\begin{equation}
\begin{aligned}
\label{stdfried2}
2\frac{\ddot S}{S}+\frac{\dot{S}^2}{S^2}+\frac{1}{2}\sigma^2-\sqrt{\frac{2}{3}}\left(\dot{\sigma}+3\frac{\dot{S}}{S}\sigma\right)=
-\frac{4}{\varphi^2_0}\,(p+p_{{\rm DE},x}),
\end{aligned}
\end{equation}
\begin{equation}
\begin{aligned}
\label{stdfried3}
2\frac{\ddot S}{S}+\frac{\dot{S}^2}{S^2}+\frac{1}{2}\sigma^2+\sqrt{\frac{1}{6}}\left(\dot{\sigma}+3\frac{\dot{S}}{S}\sigma\right)=
-\frac{4}{\varphi^2_0}\,(p+p_{{\rm DE},y}),
\end{aligned}
\end{equation}
where $\rho_{\rm DE}$ is the energy density and $p_{{\rm DE},x}$ and $p_{{\rm DE},y}$ are the principal pressures of the effective DE along the $x$-axis and the $y$- and $z$-axes, respectively, and which read
\begin{equation}
\begin{aligned}
\label{rhode}
\rho_{\rm DE}=&\frac{\varphi_0^2}{4}\bigg[ \left(\frac{4}{\varphi^2}-\frac{4}{\varphi_0^2} \right) \rho + 2\omega \frac{\dot{\varphi}^2}{\varphi^2}-6\frac{\dot{S}}{S}\frac{\dot{\varphi}}{\varphi}+2\omega M^2\bigg],
\end{aligned}
\end{equation}

\begin{equation}
\begin{aligned}
\label{pdex}
p_{{\rm DE},x}=\frac{\varphi_0^2}{4}\bigg[& \left(\frac{4}{\varphi^2}-\frac{4}{\varphi_0^2} \right)p-2\sqrt{\frac{2}{3}}\frac{\dot{\varphi}}{\varphi}\sigma\\
&+(2 \omega+2) \frac{\dot{\varphi}^2}{\varphi^2}
 +2\frac{\ddot{\varphi}}{\varphi}+4\frac{\dot{S}}{S}\frac{\dot{\varphi}}{\varphi}-2\omega M^2\bigg],
\end{aligned}
\end{equation}
\begin{equation}
\begin{aligned}
\label{pdey}
p_{{\rm DE},y}=\frac{\varphi_0^2}{4}\bigg[& \left(\frac{4}{\varphi^2}-\frac{4}{\varphi_0^2} \right)p+\sqrt{\frac{2}{3}}\frac{\dot{\varphi}}{\varphi}\sigma\\
&+(2 \omega+2) \frac{\dot{\varphi}^2}{\varphi^2}
 +2\frac{\ddot{\varphi}}{\varphi}+4\frac{\dot{S}}{S}\frac{\dot{\varphi}}{\varphi}-2\omega M^2\bigg].
\end{aligned}
\end{equation} 
We note that the Jordan field and shear scalar are coupled and they together lead to an anisotropy in the pressure of the effective DE, which may be represented by
\begin{equation}
\label{deltapde}
\Delta p_{\rm DE}=p_{{\rm DE},y}-p_{{\rm DE},x}=\sqrt{6}\frac{\varphi_0^2}{4}\frac{\dot{\varphi}}{\varphi}\sigma.
\end{equation}
This, in turn, leads to a modification in the evolution of the expansion anisotropy provided that $\varphi$ is dynamical and the space is not exactly isotropic as would be expected from a realistic cosmological model (see, e.g., \cite{Mimoso:1995ge}). We can restate this equation in terms of the cosmological gravitation coupling strength as
\begin{equation}
\label{deltapdeG}
\Delta p_{\rm DE}=-\sqrt{\frac{3}{2}}\frac{1}{8\pi G_0}\,\frac{\dot{G}}{G}\,\sigma.
\end{equation}
This effective anisotropic pressure (in the presence of shear) induced by the Jordan field in the Jordan frame is absent in the Einstein frame (see, e.g., Ref. \cite{Mimoso:1995ge}).

\section{Exact solution compatible with standard $\Lambda$CDM model}
\label{lcdmsol}
We follow the method, given in \cite{Boisseau:2010pd}, for obtaining exact cosmological solutions of a scalar-tensor gravity theory compatible with the $\Lambda$CDM model, albeit we consider LRS Bianchi type I spacetime rather than spatially flat Robertson-Walker spacetime.\footnote{We note that in the GR limit (say, $\varphi={\rm const.}$) the anisotropic model under consideration here, \eqref{eq:rho}-\eqref{eq:phi}, reduces mathematically to the standard $\Lambda$CDM+stiff matter model in the RW framework \cite{Chavanis:2014lra}. However, the role of the stiff matter with positive energy density in \cite{Chavanis:2014lra} is played by the shear scalar here, so that one can straightforwardly utilize the rich class of solutions given in \cite{Chavanis:2014lra} for the GR limit in our model. This presents a good example that one could find various solutions of the model under consideration here but yet we focus on the solution obtained by extending the method given in \cite{Boisseau:2010pd} to LRS Bianchi type I metric.} \footnote{We followed the method given in \cite{Boisseau:2010pd} for obtaining an exact cosmological solution of BD gravity theory for the expansion rate $H(z)$, with its $\Lambda$CDM model counterpart up to a large redshift, viz., for pressureless matter in a spatially flat, homogeneous and isotropic universe. A more general exact solution of the same setup was given in \cite{Uehara:1981nq} (much earlier than \cite{Boisseau:2010pd}), was nevertheless given for the scale factor in cosmic time and very complicated for extracting an exact $H(z)$ required for observational analyses.} To do so, we first re-express the set of differential equations given by Eqs. \eqref{eq:rho}-\eqref{eq:phi} using  $\frac{{\rm d}z}{{\rm d} t}=-H(1+z)$, where $H=\dot{S}/S$ is the average Hubble parameter and  $z$ is the cosmic redshift we define in terms of $S$ as $z=-1+\frac{1}{S}$. Accordingly, we re-express the \textit{energy density equation} \eqref{eq:rho} as
 \begin{equation}
  \begin{aligned}
&3H^2\bigg[1-\frac{2\omega}{3}(1+z)^2\frac{{\varphi'}^2}{\varphi^2}-2(1+z)\frac{{\varphi'}}{\varphi}\bigg]\\
&-\frac{\sigma^2}{2}-2\omega M^2=\frac{4}{\varphi^2}\rho \label{eq:rhoz},
   \end{aligned}
 \end{equation}
and obtain the \textit{pressure equation} as
  \begin{equation}
  \begin{aligned}
 &-\frac{{\rm d}H^2}{{\rm d}z}\left[1+z-(1+z)^2\frac{\varphi'}{\varphi}\right]\\
 &+3H^2\bigg\{1+(1+z)^2\bigg[\frac{2}{3}\frac{\varphi''}{\varphi}+\frac{2}{3}(1+\omega)\frac{{\varphi'}^2}{\varphi^2}\bigg]\\
 &\quad\quad\quad\quad-\frac{2}{3}(1+z)\frac{\varphi'}{\varphi}\bigg\}+\frac{\sigma^2}{2}-2\omega M^2=-\frac{4}{\varphi^2}p,\label{eq:presz}
    \end{aligned}
    \end{equation}
using Eqs. \eqref{eq:pres1} and \eqref{eq:pres2}, the \textit{shear propagation equation} as
 \begin{equation}
\frac{\sigma'}{\sigma}-\frac{3}{1+z}+2\frac{\varphi'}{\varphi}=0,
\label{shearz}
 \end{equation}
by subtracting Eq. \eqref{eq:pres1} from Eq. \eqref{eq:pres2}, and finally re-express the \textit{scalar field equation} \eqref{eq:phi} as
  \begin{equation}
  \begin{aligned}
&-\frac{{\rm d}H^2}{{\rm d}z}\left[\omega(1+z)^2\frac{\varphi'}{\varphi}+\frac{3}{2}(1+z)\right]\\
&+H^2\left[-2\omega\frac{\varphi''}{\varphi}(1+z)^2+4\omega(1+z)\frac{\varphi'}{\varphi}+6\right] \\
&+\frac{\sigma^2}{2}-2\omega M^2=0,\label{eq:phiz}
    \end{aligned}
 \end{equation}
where $'$ denotes derivative with respect to redshift (${\rm d}/{\rm d}z$).

We note that Eqs. \eqref{eq:presz} and \eqref{eq:phiz} have exactly the same mathematical form, that is, a first order linear differential equation in $H^2$, such as
\begin{align}
\label{firstordereq}
A_i(z)\frac{{\rm d}H^2}{{\rm d}z}+B_i(z)H^2+C_i(z)+D_i=0,\,\, i=1,2\, ,
\end{align}
provided that we set $p=p_{\rm m}=0$, viz., the universe is filled with only pressureless matter. We next note that constants $D_1=D_2=-2\omega M^2$, and that the shear scalar $\sigma^2$ is the term that differs in our system of equations from the one given in \cite{Boisseau:2010pd}. Fortunately, it contributes to \eqref{eq:presz} and \eqref{eq:phiz} in the same way, namely, as $C_1(z)=C_2(z)=\frac{\sigma^2}{2}$. In accordance with these points, we assume $A_1(z)=A_2(z)$ in \eqref{eq:presz} and \eqref{eq:phiz} and then solve for the rate of change of Jordan field in $z$ as
\begin{align}
\frac{\varphi'}{\varphi}=-\frac{1}{2(1+\omega)}\frac{1}{1+z},
\label{eq:Fgeneral}
\end{align}
which in turn renders the coefficients of $H^2$ identical, i.e, $B_1(z)=B_2(z)$, as well. Thus, integrating \eqref{eq:Fgeneral}, it turns out that the solution of the Jordan field is
\begin{align}
\varphi^2=\varphi_0^2(1+z)^{-\frac{1}{1+\omega}},
\label{eq:phii}
\end{align}
where $\varphi_0=\varphi(z=0)$ is the present time value of the Jordan field, and which in turn gives
\begin{align}
G=G_{0} (1+z)^{\frac{1}{1+\omega}}.
\label{eq:Geff}
\end{align}
The integration of the shear propagation equation given in Eq. \eqref{shearz} gives
\begin{align}
\sigma^2=\sigma_0^2(1+z)^6\left(\frac{\varphi}{\varphi_0}\right)^{-4},
\end{align}
where $\sigma^2_0=\sigma^2(z=0)$ is the present time value of the shear scalar, leading, together with \eqref{eq:phii}, to
\begin{align}
\sigma^2=\sigma_0^2(1+z)^{6+\frac{2}{1+\omega}},
\label{eq:sigma}
\end{align}
whereas it is $\sigma_{\rm GR}^2 \propto (1+z)^6$ in GR (corresponding to the $| \omega |  \rightarrow\infty$ limit of the BD gravity), and thereby the energy density corresponding to the shear scalar can be defined as 
\begin{align}
\rho_{\sigma^2}\equiv\frac{\sigma^2}{2}\frac{\varphi_0^2}{4}=\frac{\sigma^2_0}{2}\frac{\varphi_0^2}{4}(1+z)^{6+\frac{2}{1+\omega}},
\label{eq:sigmarho}
\end{align}
by considering the present time value of the Jordan field, i.e., of the cosmological gravitational coupling strength. Because BD theory \cite{Brans:1961sx} satisfies the local conservation of EMT,  the energy density of the pressureless matter can immediately be written as
\begin{equation}
\rho_{\rm m}=\rho_{\rm m,0}(1+z)^3,
\label{eqn:rhozn}
\end{equation}
which is exactly the same as the one for the pressureless matter in the standard $\Lambda$CDM model. Finally using $\varphi^2$, $\sigma^2$ and $\rho_{\rm m}$ from Eqs. \eqref{eq:phii}, \eqref{eq:sigma} and \eqref{eqn:rhozn}, respectively, in Eq. \eqref{eq:rhoz} we reach the following modified anisotropic Friedmann equation for BD theory in this solution
\begin{equation}
\begin{aligned}
H^2=&\frac{4}{3}\frac{\gamma}{\varphi_0^2}\big[\rho_{\rm M}+\rho_{{\rm m},0}(1+z)^{3+\frac{1}{1+\omega}}+\rho_{\sigma^2,0}(1+z)^{6+\frac{2}{1+\omega}}\big],
\label{eq:fried1}
\end{aligned}
\end{equation}
where
\begin{equation}
\label{gammadef}
\gamma=\frac{6(1+\omega)^2}{(3\omega+4)(2\omega+3)},
\end{equation}
and the energy densities corresponding to the mass of the Jordan field and to the expansion anisotropy today read, respectively,
\begin{equation}
\begin{aligned}
\label{rhoss}
\rho_{\rm M}=2M^2\omega\frac{\varphi_0^2}{4}\quad \textnormal{and} \quad \rho_{\sigma^2,0}=\frac{\sigma_0^2}{2}\frac{\varphi_0^2}{4}. 
  \end{aligned}
  \end{equation}
One may check for consistency that we recover the $H(z)$ of the standard $\Lambda$CDM model when we set $\omega\rightarrow\infty$ (viz., in the GR limit of BD gravity giving $\varphi\rightarrow {\rm constant}$) and $\rho_{\sigma^2,0}=0$ (viz., isotropic expansion as in the RW spacetime metric). We also note that $\rho_{\rm M}$ can be regularized to give a finite positive value consistent with the observations by setting  $M^2\omega={\rm constant}$ so that $M^2\rightarrow 0$ as $\omega\rightarrow\infty$ such that the term $M^2\omega$ arising from their multiplication would always remain unaltered and finite.

\subsection{Inclusion of radiation}
\label{radsol}

The solution of the BD extension of standard $\Lambda$CDM model, for either isotropic or anisotropic spatially flat spacetimes, can be achieved by using the method given in Ref. \cite{Boisseau:2010pd} provided that $p=0$, viz., the Universe is filled with only pressureless matter. It covers also the epoch of the Universe when the contributions from the pressureless matter in \eqref{eq:fried1}, namely, $\frac{4}{\varphi^2}\rho_{\rm m}\propto (1+z)^{3+\frac{1}{1+\omega}}$, is dominant over the effective cosmological constant $2\omega M^2$ and one may check that for this epoch, neglecting anisotropy, it reduces to the well known solution with $S\propto t^{\frac{2+2\omega}{4+3\omega}}$ and $\varphi^2\propto t^{\frac{2}{4+3\omega}}$ (see, e.g., \cite{Clifton:2005xr} and references therein). We could not obtain analytically an exact cosmological solution -- other than the trivial solution with $\varphi=\rm const.$-- using the same method when we include relativistic source, $p_{\rm r}=\frac{\rho_{\rm r}}{3}$. Fortunately the effects of the relativistic source on the background dynamics can be neglected in the late Universe. Namely, because local energy conservation holds in BD gravity in the Jordan frame, we have $\rho_{\rm r}\propto (1+z)^4$ and $\rho_{\rm m}\propto (1+z)^3$, implying that the effects of the relativistic source on the dynamics of the Universe will be significant for large redshift values, in particular, for the redshift values larger than the matter-radiation equality redshift, $z_{\rm eq}=-1+\frac{\rho_{\rm m,0}}{\rho_{\rm r,0}}\sim 3380$ \cite{Aghanim:2018eyx}, as in the standard cosmology \cite{Liddle:1998ij}. The attractor solution of the BD gravity for radiation domination is well known that it is exactly the GR solution, i.e., $\varphi= {\rm const.}$ \cite{Nariai69,Gurevich73,Barrow:1993nt,Liddle:1998ij,Clifton:2005xr}. Indeed, in general, BD cosmologies have exact solutions which show that they are dominated by the Jordan field at early times and by the perfect fluid matter sources at \textit{late times}, which here can be considered as the period all the way down from matter-radiation equality to the earlier times covering the times relevant to Big Bang Nucleosynthesis (BBN). Hence, for $z>z_{\rm eq}$, we have $\varphi= \varphi_1= {\rm const.}$, implying $G=G_1= {\rm const.}$ as long as anisotropy is negligible. Note that $\varphi_1$ (or $G_1$) will be different than $\varphi_0$ (or $G_0$) as a consequence of the fact that the Jordan field has been evolving approximately in accordance with \eqref{eq:phii} for $z<z_{\rm eq}$, i.e., between the matter-radiation equality and today. Accordingly, our model will more approach a general relativistic cosmology as the radiation dominates over pressureless matter as we go to higher redshifts, yet the most part of period of the Universe during which $z<z_{\rm eq}$ the Jordan field evolves approximately in accordance with \eqref{eq:phii} and thereby we can estimate that
\begin{equation}
\begin{aligned}
\varphi_1\sim\varphi_0(1+z_{\rm eq})^{-\frac{1}{2+2\omega}}\quad&\textnormal{at}\quad z\sim z_{\rm eq} \\
\quad\quad\quad\quad \textnormal{as well as} \quad \varphi\sim \varphi_1\quad&\textnormal{for}\quad z\gtrsim z_{\rm eq},
\label{eqn:freezingphi}
\end{aligned}
\end{equation}
which implies
\begin{align}
\label{G1G0rel}
G_1\sim G_0(1+z_{\rm eq})^{\frac{1}{1+\omega}} \quad\textnormal{for}\quad z\gtrsim z_{\rm eq},
\end{align}
where $G_1$ is then approximately the cosmological gravitational coupling strength for $z>z_{\rm eq}$ (see \cite{Clifton:2005xr} for further details). Thus, for $\omega>0$, the cosmological gravitational coupling strength will remain constant (provided that anisotropy is negligible) during the radiation dominated epoch as just like in the standard cosmology, but will be larger than its present time value, e.g., $G_1/G_0\sim 3380^{\frac{1}{1+\omega}}$. This implies an enhanced expansion rate of the Universe during this epoch, which in turn will affect, for instance, the primordial element abundances. We shall further discuss this point in Sect.~\ref{sec:varGandBBN}. We next see from \eqref{shearz} that we approximately have $\sigma^2\propto\sigma_0^2(1+z)^6$ like in GR in the presence of isotropic perfect fluid, since we have $\varphi\sim{\rm constant}$ when the radiation is dominant over pressureless matter and expansion anisotropy is still insignificant. As we go to even larger redshift values, say, $z\gg z_{\rm BBN}$, expansion anisotropy will dominate over the terms due to the radiation and hence we obtain once again $\rho_{\sigma^2}\propto (1+z)^{6+\frac{2}{1+\omega}}$ (which can be obtained simply by setting $\rho_{\rm M}=0$ and $\rho_{\sigma^2,0}=0$ in our solution).

In the light of the above discussion, although we do not have exact analytical solution when we include the radiation as well, we can find out the contribution from the radiation. First of all, it is obvious that during radiation domination the Jordan field is constant, then the inclusion of radiation to the model will in general slow-down the Jordan field, namely, the model in general will deviate less from GR. Of course, that effect will be insignificant for a viable and realistic cosmological model [namely, cosmological model satisfying, roughly, $\rho_{\sigma^2,0}\ll\rho_{\rm r,0}\ll\rho_{\rm m,0}\sim \rho_{\rm M}$ for $\omega\gtrsim 10$ (see Sect.~\ref{sub1preliminary} for this choice on $\omega$)] for $z< z_{\rm eq}$. Thus, we can write
\begin{equation}
\begin{aligned}
H^2\approx\frac{4}{3}\frac{\gamma}{\varphi_0^2}\bigg[&\rho_{\rm M}+\rho_{{\rm m},0}(1+z)^{3+\frac{1}{1+\omega}}+\rho_{{\rm r},0}(1+z)^{4+\frac{1}{1+\omega}}\\
&+\rho_{\sigma^2,0}(1+z)^{6+\frac{2}{1+\omega}}\bigg]\quad\textnormal{for}\quad z<z_{\rm eq},
\label{eq:fried1zlate}
\end{aligned}
\end{equation}
which can be \textit{safely} used all the way to the recombination redshift (CMB release), such that it is well known that the Universe should have transited from radiation to matter dominated era when $z\sim 3380$, and the recombination that leads to photon decoupling should have taken place when $z\sim 1100$, at which  the pressureless matter is still dominant over radiation. Therefore, we note that $H(z)$ given in \eqref{eq:fried1zlate} can safely be used for constraining the model by using CMB radiation data as well.

For the times before the matter-radiation equality, setting $\varphi$ constant relying on its attractor behavior during the radiation-dominated phase, we can write 
\begin{equation}
\begin{aligned}
H^2\approx\frac{4}{3}\frac{\gamma}{\varphi_1^2}\left[\rho_{{\rm r},0}(1+z)^{4}+\rho_{\sigma^2,1}(1+z)^{6}\right]\\
\quad \textnormal{for}\quad z_{\rm eq}<z<z_{\rm eq,\sigma^2,r}.
\label{eq:fried1zearly}
\end{aligned}
\end{equation}
Note that, for consistency between \eqref{eq:fried1zlate} and \eqref{eq:fried1zearly}, here in \eqref{eq:fried1zearly} we have $\varphi_1\sim\varphi_0(1+z_{\rm eq})^{-\frac{1}{2+2\omega}}$ in contrast to $\varphi_{0}$ in \eqref{eq:fried1zlate}, in accordance with the above discussions leading to \eqref{eqn:freezingphi}, and similarly $\rho_{\sigma^2,1}\sim\rho_{\sigma^2,0}(1+z_{\rm eq})^{\frac{1}{1+\omega}}$. Using \eqref{eq:fried1zearly}, we estimate (at least roughly) the radiation-expansion anisotropy equality redshift as $z_{\rm eq,\sigma^2,r}\sim -1+ \sqrt{\frac{\rho_{\rm r,0}}{\rho_{\sigma^2,1}}}$, which can also be written as follows;
\begin{equation}
\begin{aligned}
\label{eqn:eqsigmar}
z_{\rm eq,\sigma^2,r}&\sim-1+ \sqrt{\frac{\rho_{\rm r,0}}{\rho_{\sigma^2,0}}\left(\frac{\rho_{\rm m,0}}{\rho_{\rm r,0}}\right)^{-\frac{1}{1+\omega}}}.
\end{aligned}
\end{equation}
Avoiding expansion anisotropy from spoiling the physical processes relevant to BBN will then roughly require $z_{\rm eq,\sigma^2,r}>z_{\rm BBN}$. Typically, in a viable cosmological model, we expect the matter-radiation equality redshift to be $z_{\rm eq}=-1+\frac{\rho_{\rm m,0}}{\rho_{\rm r,0}}\sim 3380$, the BBN to take place at $z_{\rm BBN}\sim 3\times 10^{8}$ and $\Omega_{\rm r,0}\sim 10^{-4}$. Accordingly, taking $z_{\rm eq,\sigma^2,r}> 3\times 10^{8}$, we obtain $\frac{\rho_{\sigma^2, 0}}{\rho_{\rm r, 0}}=\frac{\Omega_{\sigma^2, 0}}{\Omega_{\rm r, 0}}\lesssim 10^{-17-\frac{4}{1+\omega}}$, namely, the following stringent constraint on the density parameter of the expansion anisotropy today:
\begin{equation}
\label{omsigma0constraint}
\Omega_{\sigma^2,0}\lesssim 10^{-21-\frac{4}{1+\omega}},
\end{equation}
which reduces to $\Omega_{\sigma^2,0}\sim 10^{-21}$ in the GR limit ($\omega\rightarrow\infty$). It is noteworthy that, in the GR limit, at which anisotropy in the pressure also disappears, our estimation on the expansion anisotropy for today is already of the same order of magnitude with the ones given in \cite{Saadeh:2016sak} from a general test of isotropy using CMB temperature and polarization data from Planck and in \cite{Barrow:1976rda} from primordial nucleosynthesis, both of them consider GR and only isotropic fluids. This also shows the reliability of our estimation. We note that, for the case $\omega>0$ we consider in this study, BD theory leads to more stringent constraints on the expansion anisotropy (e.g., $\Omega_{\sigma^2,0}\lesssim 10^{-25}$ for $\omega=0$), while the case $\omega<-1$, which we do not consider in the present study, can relax it. For instance, Campanelli et al. \cite{Campanelli:2010zx} have put model-independent upper bounds on the expansion anisotropy, from type Ia SNe observations, in the late Universe, namely, for $z\lesssim 1.6$, as $\Omega_{\sigma^2,0}\lesssim 10^{-4}$ (see also \cite{Wang:2017ezt} for a recent work). We note that this value is achieved upon choosing $\omega=-\frac{21}{17}$ in \eqref{omsigma0constraint}, which can be done without worrying for theoretical issues associated with $\omega<-\frac{3}{2}$, but this will cost large deviations from GR (viz. rapidly varying strength of the gravitational coupling) as well as the standard $\Lambda$CDM. We shall further discuss possible modifications on the expansion anisotropy for different values/ranges of $\omega$ and their comparisons with the standard $\Lambda$CDM model in Sect.~\ref{anidiscuss}.

Finally, for the times when the expansion anisotropy dominates over radiation (and hence obviously on pressureless matter), we can write 
\begin{equation}
H^2\propto (1+z)^{6+\frac{2}{1+\omega}}\quad\textnormal{for}\quad z>z_{\rm eq,\sigma^2,r},
\label{eq:friedani}
\end{equation}
from our solution (see, e.g., Refs. \cite{CervantesCota:1999tx,CervantesCota:2000wc,Mimoso:1995ge,Kamenshchik:2017ojc} for anisotropy dominated universes in the context of scalar-tensor theories). This epoch however is out of interest in the present work.

\subsection{Effective anisotropic dark energy} 
\label{sec:effde}
The discussions in the previous subsections reveal that the massive BD gravity modification on the gravity sector, manifests itself when radiation becomes subdominant, say, in the relatively late Universe, that is, switching from GR+$\Lambda$ to the massive BD introduces corrections on the redshift dependencies of both the average expansion rate and expansion anisotropy. Although we consider the presence of radiation when we constrain the model using the cosmological data, while we are discussing the effective DE in what follows, we shall neglect its presence due to the following reasons: (i) Its effect on the effective DE will obviously be negligible in the late universe. (ii) When it is dominant the Jordan field $\varphi$ freezes, implying the BD gravity mimics GR, and thereby the effective DE will mimic nothing but $\Lambda$ (not to mention that DE would be subdominant at this epoch). (iii) A practical reason, such that we don't have explicit analytical solution when we include radiation into the model. On the other hand, although we expect anisotropy to be negligible (even with respect to radiation) in the late universe, we keep expansion anisotropy due to the following reasons: (i) To see its effect on the effective DE explicitly. (ii) It leads to an anisotropy in the pressure of the effective DE, which in turn modifies the redshift dependency of the expansion anisotropy w.r.t. GR ($|\omega|\rightarrow\infty$). (iii) In relevance with (ii), its modified redshift dependency w.r.t. the one in GR (provided that it can be detected) would be a strong signal in favor of modified gravities (such as BD theory), rather than the presence of a DE (such as $\Lambda$, scalar fields) ingredient of the Universe assuming that GR is the true theory of gravity.

 We implement the solution given in Sect.~\ref{lcdmsol} and accordingly we re-write \eqref{stdfried} as follows:
\begin{equation}
\begin{aligned}
\label{stdfriedwithrho}
3H^2=\frac{4}{\varphi^2_0}\,\left(\rho_{\rm m}+\rho_{\rm DE}+\rho_{\sigma^2}\right),
\end{aligned}
\end{equation}
where the energy density of the effective DE, $\rho_{\rm DE}$, reads
\begin{equation}
\label{rhodesol}
\rho_{\rm DE}=\gamma \rho_{\rm M}+\rho_{\rm m}\left[\gamma\left(\frac{\rho_{\rm m}}{\rho_{\rm m,0}}\right)^{\frac{1}{3(1+\omega)}}-1\right]+\rho_{\sigma^2}(\gamma-1),
\end{equation}
which gives
\begin{equation}
\begin{aligned}
\label{rhode0comp}
\rho_{\rm DE,0}=&\gamma \rho_{\rm M}+\rho_{\rm m,0}\left(\gamma-1\right)+\rho_{\sigma^2,0}(\gamma-1)
\end{aligned}
\end{equation}
 for $z=0$. Dividing \eqref{stdfriedwithrho} by $3H_0^2$, we obtain
\begin{equation}
\begin{aligned}
\label{hsquared}
\frac{H^2}{H_0^2}=\Omega_{\rm DE,0}f(z)+\Omega_{\rm m,0}(1+z)^3+\Omega_{\sigma^2,0}(1+z)^{6+\frac{2}{1+\omega}},
\end{aligned}
\end{equation}
where we define $f(z)=\frac{\rho_{\rm DE}}{\rho_{\rm DE,0}}$, reading
\begin{align}
\label{Omde}
f(z)=\,1+\frac{\Omega_{\sigma^2,0}}{\Omega_{\rm DE,0}}\left(\gamma-1\right)\left[(1+z)^{6+\frac{2}{1+\omega}}-1\right]\nonumber \\
+\frac{\Omega_{\rm m,0}}{\Omega_{\rm DE,0}}\left\{(1+z)^3\left[\gamma(1+z)^{\frac{1}{1+\omega}}-1\right]+1-\gamma\right\},
\end{align}
which gives $f(z=0)=1$ consistently with $\Omega_{\rm m,0}+\Omega_{\rm DE,0} + \Omega_{\sigma^2,0}=1$. Note that we define the present time values of the density parameters as $\Omega_{i,0}=\rho_{i,0}/\rho_{\rm c,0}$, where $\rho_{\rm c,0}=3H_0^2\varphi_0^2/4$ is the present time value of the critical energy density, and that we eliminate $\Omega_{\rm M,0}$, that appears in $f(z)$ originally, using the relation
\begin{equation}
\begin{aligned} 
\label{omegaMdel}
\Omega_{{\rm M},0}=\gamma^{-1}\left[\Omega_{{\rm DE},0}-(\gamma-1)(\Omega_{\rm m,0}+\Omega_{\sigma^2,0})\right],
\end{aligned}
\end{equation}
which can easily be obtained from \eqref{rhodesol}. Using \eqref{omegaMdel} along with $\Omega_{\rm m,0}+\Omega_{\sigma^2,0}=1-\Omega_{\rm DE,0}$, we see that the present time density parameter of the effective DE is determined by the Jordan field's mass and the BD parameter $\omega$ [viz., $\gamma(\omega)$] as \footnote{It may be interesting to note here that for the case $M=0$ (i.e., the massless Jordan field case), $\Omega_{{\rm DE},0}$ is a function of $\omega$ only, and if we set $\Omega_{{\rm DE},0}=0.7$ we find that $\omega=-\frac{5}{4}$ from \eqref{gammaMass} and ${\bar w}_{\rm DE,0}\approx-1.29$ from \eqref{wde0} assuming the contribution from the expansion anisotropy is insignificant (presence of expansion anisotropy shifts it to even larger negative values). This ($\omega=-\frac{5}{4}$), in turn, implies from \eqref{eq:sigma} that the shear scalar (viz., expansion anisotropy) contributes to the modified Einstein field equations interpreted in accordance with \eqref{einsteinian} like a phantom source with an effective equation of state as $p=-\frac{5}{3}\rho$.}
\begin{align}
\label{gammaMass}
\Omega_{{\rm DE},0}=\omega\frac{2}{3}\left(\frac{M}{H_0}\right)^2+\frac{\gamma-1}{\gamma}.
\end{align}

 We obtain --upon straightforward arrangements in the equations obtained by using \eqref{eq:fried1}, \eqref{omegaMdel} in the equations obtained by substituting \eqref{eq:phii}, \eqref{gammadef}, \eqref{rhoss} in \eqref{pdex}, \eqref{pdey}-- the principal EoS parameters of the effective DE along the $x$-axis and $y$- and $z$-axes in terms of the present day values of the density parameters as follows, respectively,

\begin{align}
\label{wx1}
w_{{\rm DE},x}\equiv&\frac{p_{{\rm DE},x}}{\rho_{\rm DE}}=-1+\frac{1}{f(z)}\bigg[\frac{\Omega_{\rm m,0}}{\Omega_{\rm DE,0}}\frac{2\omega+2}{2\omega+3}(1+z)^{3+\frac{1}{1+\omega}}\nonumber \\
&-\frac{\Omega_{\rm m,0}}{\Omega_{\rm DE,0}}(1+z)^{3}-\frac{2}{2\omega+3}\frac{\Omega_{\sigma^2,0}}{\Omega_{\rm DE,0}}(1+z)^{6+\frac{2}{1+\omega}}\nonumber \\
&-\frac{2}{3(1+\omega)}\sqrt{\frac{\Omega_{\sigma^2,0}}{\Omega_{\rm DE,0}}} (1+z)^{3+\frac{1}{1+\omega}}\nonumber \\
&\sqrt{1+\frac{\Omega_{\rm m,0}}{\Omega_{\rm DE,0}}(1+z)^{3}+\frac{\Omega_{\sigma^2,0}}{\Omega_{\rm DE,0}}(1+z)^{6+\frac{2}{1+\omega}}} \bigg],
\end{align}
\begin{align}
\label{wy1}
w_{{\rm DE},y}\equiv&\frac{p_{{\rm DE},y}}{\rho_{\rm DE}}=-1+\frac{1}{f(z)}\bigg[\frac{\Omega_{\rm m,0}}{\Omega_{\rm DE,0}}\frac{2\omega+2}{2\omega+3}(1+z)^{3+\frac{1}{1+\omega}}\nonumber \\
&-\frac{\Omega_{\rm m,0}}{\Omega_{\rm DE,0}}(1+z)^{3}-\frac{2}{2\omega+3}\frac{\Omega_{\sigma^2,0}}{\Omega_{\rm DE,0}}(1+z)^{6+\frac{2}{1+\omega}}\nonumber \\
&+\frac{1}{3(1+\omega)}\sqrt{\frac{\Omega_{\sigma^2,0}}{\Omega_{\rm DE,0}}}(1+z)^{3+\frac{1}{1+\omega}}\nonumber \\
&\sqrt{1+\frac{\Omega_{\rm m,0}}{\Omega_{\rm DE,0}}(1+z)^{3}+\frac{\Omega_{\sigma^2,0}}{\Omega_{\rm DE,0}}(1+z)^{6+\frac{2}{1+\omega}}} \bigg],
\end{align}

which give
\begin{align}
w_{{\rm DE},x,0}=-1-\frac{\Omega_{\rm m,0}+2\,\Omega_{\sigma^2,0}}{(2\omega+3)\Omega_{\rm DE,0}}
-\frac{2\sqrt{\Omega_{\sigma^2,0}}}{3(1+\omega)\Omega_{\rm DE,0}},\\
w_{{\rm DE},y,0}=-1-\frac{\Omega_{\rm m,0}+2\,\Omega_{\sigma^2,0}}{(2\omega+3)\Omega_{\rm DE,0}}
+\frac{\sqrt{\Omega_{\sigma^2,0}}}{3(1+\omega)\Omega_{\rm DE,0}}
\end{align}
for $z=0$.

The anisotropy of the EoS of the effective DE may be given as the difference between its principal EoS parameters ($\Delta w_{\rm DE}=w_{{\rm DE},y}-w_{{\rm DE},x}$) as
\begin{equation}
\begin{aligned}
\label{deltaw1}
\Delta w_{\rm DE}=&\frac{1}{(1+\omega)}\frac{(1+z)^{3+\frac{1}{1+\omega}}}{f(z)} \sqrt{\frac{\Omega_{\sigma^2,0}}{\Omega_{\rm DE,0}}} \\
&\times\sqrt{1+\frac{\Omega_{\rm m,0}}{\Omega_{\rm DE,0}}(1+z)^{3}+\frac{\Omega_{\sigma^2,0}}{\Omega_{\rm DE,0}}(1+z)^{6+\frac{2}{1+\omega}}}.
\end{aligned}
\end{equation}
Accordingly, we have
\begin{equation}
\label{deltaw1now}
\Delta w_{\rm DE,0}=\frac{1}{1+\omega}\,\frac{\sqrt{\Omega_{\sigma^2,0}}}{\Omega_{\rm DE,0}}\quad\textnormal{for}\quad z=0,
\end{equation}
which is approximately equal to $\frac{1}{1+\omega}\sqrt{\frac{\Omega_{\sigma^2,0}}{\Omega_{\rm DE,0}}}$, provided that the effective DE is dominant today ($z=0$). Additionally, we can write for dust domination
\begin{equation}
\label{eqn:anisoEoSdust}
\Delta w_{\rm DE}\sim\frac{1}{(1+\omega)\gamma}\sqrt{\frac{\Omega_{\sigma^2,0}}{\Omega_{\rm m,0}}}\,(1+z)^{\frac{3}{2}},
\end{equation}
which is positive definite as $\omega>0$, and for expansion anisotropy domination
\begin{equation}
\label{eqn:anisoEoSaniso}
\Delta w_{\rm DE}\sim-\frac{1}{1+\omega}\,\frac{1}{1-\gamma},
\end{equation}
which depends only on $\omega$, is negative definite as $\omega>0$, monotonic in $\omega>0$ and takes values as $\Delta w_{\rm DE}=\{-2,-1.2\}$ for $\omega=\{0,\infty\}$.

We finally, using $\dot{\rho}_{\rm DE}+3H\rho_{\rm DE}(1+{\bar w}_{\rm DE})=0$, define a mean/volumetric effective EoS parameter for the effective DE as
\begin{equation}
\begin{aligned}
\label{averageeos}
{\bar w}_{\rm DE}=-1+\frac{1+z}{3}\frac{\rho_{{\rm DE}}'}{\rho_{\rm DE}}=-1+\frac{1+z}{3}\frac{f'(z)}{f(z)},
\end{aligned}
\end{equation}
which explicitly reads
\begin{widetext}
\begin{equation}
\begin{aligned}
\label{wdedef}
{\bar w}_{\rm DE}=&-1+\frac{\Omega_{{\rm m},0}(1+z)^3\left[\gamma\frac{3\omega+4}{3(1+\omega)}(1+z)^{\frac{1}{1+\omega}}-1\right]+\Omega_{{\sigma^2},0}\frac{2(\gamma-1)(3\omega+4)}{3(1+\omega)}(1+z)^{6+\frac{2}{1+\omega}}}{\Omega_{\rm DE,0}+\Omega_{\rm m,0} \left\{(1+z)^3\left[\gamma(1+z)^{\frac{1}{1+\omega}}-1\right]+1-\gamma\right\}+\Omega_{\sigma^2,0}(\gamma-1)\left[(1+z)^{6+\frac{2}{1+\omega}}-1\right]}, 
\end{aligned}
\end{equation}
\end{widetext}
and
\begin{equation}
\begin{aligned}
\label{wde0}
{\bar w}_{\rm DE,0}=&-1-\frac{1}{2\omega+3}\,\frac{\Omega_{{\rm m},0}}{\Omega_{\rm DE,0}}-\frac{2(5\omega+6)}{3(2\omega+3)(\omega+1)}\,\frac{\Omega_{{\rm \sigma^2},0}}{\Omega_{\rm DE,0}}
\end{aligned}
\end{equation}
for $z=0$. We note that, given $\Omega_{\rm DE,0}>0$ and $\Omega_{\rm m,0}$ and $\Omega_{{\sigma^2},0}$ are already positive definite, for $\omega>0$, the contributions from both the pressureless matter and the expansion anisotropy to the present value of the mean EoS of the effective DE, ${\bar w}_{\rm DE,0}$, are negative, i.e., anisotropic BD extension of the standard $\Lambda$CDM model predicts phantom like effective DE in the present time Universe.
\bigskip

\subsection{Preliminary investigations}
\label{preliminary}

\subsubsection{Features of effective anisotropic dark energy}
\label{sub1preliminary}

In this section, before the observational analysis, we discuss some of the features of the model, in particular those corresponding to the effective DE and the evolution of the expansion anisotropy. To do so, we start with justifying the positivity assumption on $\omega$ in our study by making use of one of the important parameters about the kinematics of the Universe, the deceleration parameter. We define the mean/volumetric deceleration parameter in terms of the average Hubble parameter as $q=-1+~\frac{H'}{H}(1+z)$. This reads, as we expect in a viable cosmology $\Omega_{{\sigma^2},0}<\Omega_{\rm r,0}\ll \Omega_{{\rm m},0}$,
\begin{equation}
\begin{aligned}
q\approx-1+\frac{4+3\omega}{2+2\omega}\,\frac{\Omega_{{\rm m},0}(1+z)^{3+\frac{1}{1+\omega}}}{1+\gamma\Omega_{{\rm m},0}\left[(1+z)^{3+\frac{1}{1+\omega}}-1\right]}
\end{aligned}
\end{equation}
for $z\sim0$, and
\begin{equation}
\begin{aligned}
\label{qM}
q_0\approx-1+\frac{4+3\omega}{2+2\omega}\Omega_{{\rm m},0},
\end{aligned}
\end{equation}
for $z=0$ today. We confine the present study to small deviations from the standard $\Lambda$CDM model and hence it is reasonable to expect, in accordance with the most recent Planck results \cite{Aghanim:2018eyx}, that $ \Omega_{{\rm m},0}\sim 0.3$ leading to $q_0\sim -0.55$ from $q=-1+\frac{3}{2}\Omega_{{\rm m}}$ in the standard $\Lambda$CDM, which in turn implies from \eqref{qM} that we should have $|\omega| \gtrsim 10$, which leads to $ \omega \gtrsim 10$ since have already assumed $\omega\geq 0$.\footnote{Measured current values and fitting results of $q_0$ obtained from different models are given in \cite{Gomez-Valent:2018hwc,Akarsu:2013lya}, see also \cite{Gomez-Valent:2018hwc} for both the history of the accurate estimations of $q$ and recent $q$ parametrizations  without focusing in any concrete cosmological model, using cosmography.} Unless otherwise, for instance, if $\omega=0$, then $q_0\sim -0.4$, which could be decreased to a reasonable value, viz., $q_0 \sim -0.55$, by decreasing $\Omega_{{\rm m},0}$ considerably, namely $\Omega_{{\rm m},0}\sim 0.22$, which implies very large deviations from the standard $\Lambda$CDM. Therefore, our assumption $\omega\geq 0$ already allows large deviations from the standard $\Lambda$CDM, in other words, we do not oversimplify the model but yet basically avoid the cases most likely non-viable that would lead to extended discussions since the model exhibits various complicated dynamics particularly at small negative values of $\omega$, namely, when $\omega\sim-1$. Hence, in this section, we shall depict some key/interesting features of the anisotropic BD model and its deviation from the standard $\Lambda$CDM model assuming $\omega\geq0$.

First of all, one may check easily that, similar to the standard $\Lambda$CDM, its anisotropic BD extension asymptotically approaches to de Sitter Universe in the far future,
\begin{equation}
\begin{aligned}
H^2\rightarrow\frac{4}{3}\frac{\gamma}{\varphi_0^2}\rho_{\rm M}=\frac{2\gamma}{3}\omega M^2,\,\,\, \rho_{\sigma^2}\rightarrow 0\,\,\,\textnormal{and} \,\,\,
 \rho_{\rm m}\rightarrow 0 \\ \,\,\,\textnormal{for}\,\,\, z\rightarrow -1,
\end{aligned}
\end{equation}
and accordingly, at this limit, the effective DE mimics exactly the cosmological constant as
\begin{equation}
\label{rhodeM1}
\rho_{\rm DE}\rightarrow \gamma \rho_{\rm M},\,\,\, \bar w_{\rm DE}\rightarrow -1\quad\textnormal{and}\quad \Delta w_{\rm DE}\rightarrow 0.
\end{equation}

At low redshifts, viz., at which $\rho_{\rm m}\simeq\rho_{{\rm m},0}$ and $\rho_{\sigma^2}\ll\rho_{\rm m}$, the effective DE can approximately be described by
\begin{equation}
\begin{aligned}
\label{rhode1}
&\rho_{\rm DE}\simeq \gamma \rho_{\rm M}+(\gamma-1)\rho_{\rm m }\quad\textnormal{and}\quad\\
&\bar w_{\rm DE}\lesssim -1+\frac{(\gamma-1)\rho_{\rm m }}{\gamma \rho_{\rm M}+(\gamma-1)\rho_{\rm m }}\quad\textnormal{for}\quad z\simeq0.
\end{aligned}
\end{equation}

Here, we first note that the non-negativity of the effective DE today ($z=0$), $\rho_{\rm DE,0}\geq0$, leads to a natural lower bound on the value of the energy density corresponding to the effective cosmological constant (viz., $\rho_{\rm M}$) as $\rho_{\rm M}\geq\left(\frac{1}{\gamma}-1\right)\rho_{{\rm m},0}$, which allows zero lower bound only at the GR limit, i.e., at $\omega\rightarrow\infty$ (viz., $\gamma\rightarrow1$). We next note that $\gamma$ takes values within the range $\frac{1}{2}\leq\gamma< 1$ for $\omega\geq0$ and hence the coefficient of $\rho_{\rm m}$ is negative definite, $\gamma-1< 0$, which implies that, because $\rho_{\rm m}$ decreases as the Universe expands, $\rho_{\rm DE}$ will increase as the Universe expands at $z\sim 0$, i.e., the effective DE will behave like a phantom fluid ($\bar{w}_{\rm DE}\lesssim -1$) at $z\sim 0$ [see \eqref{wde0} for $\bar{w}_{\rm DE}(z=0)$]. Namely, we see from \eqref{rhodesol} or \eqref{wdedef} that given $\rho_{\rm m}=\rho_{\rm m,0}(1+z)^3$, the effective DE passes below the phantom-divide-line (PDL, $w_{\rm DE}=-1$) at
\begin{equation}
\label{rhomsignz}
 z=z_{\rm PDL}\equiv\left(\frac{2\omega+3}{2\omega+2}\right)^{1+\omega}-1, 
\end{equation}
and afterwards stays there forever, i.e., $\bar w_{\rm DE}<-1$ for $z<z_{\rm PDL}$. It is worth noting that the BD gravity predicts almost a specific redshift for the PDL crossing, such that
\begin{equation}
\frac{1}{2}\leq z_{\rm PDL}\leq e^{\frac{1}{2}}-1=0.65\quad \textnormal{for}\quad \omega\geq 0, 
\end{equation}
and $0.63\leq z_{\rm PDL}<0.65$ for $\omega\geq10$ (see  \cite{Boisseau:2010pd} for a similar result). This is indeed an interesting result, such that $z_{\rm PDL}\sim 0.6$ can be taken as the signature of BD gravity. In other words, observations suggesting the presence of a DE source passing below PDL at $z\sim0.6$ with high confidence would imply a strong reason for favoring the BD gravity over GR, or vice versa. During this period, $\Delta w_{\rm DE}$ can approximately be given by \eqref{deltaw1now} and we note that the effective DE will be almost isotropic, $\Delta w_{\rm DE}\approx0$, for  $\omega>0$ since $\Omega_{\sigma^2,0}\ll\Omega_{\rm DE,0}$ in a realistic cosmology.

At moderate redshifts (e.g., $z_{\rm PDL} \ll z \lesssim z_{\rm eq}$), namely, when the contribution from the pressureless matter to the EoS parameter of the effective DE given in \eqref{wdedef} becomes significant but those from the expansion anisotropy and the mass of the Jordan field are not, the mean/volumetric EoS parameter of effective DE yields a plateau where
\begin{equation}
\begin{aligned}
\label{wdedefmod}
\rho_{\rm DE}\sim\rho_{\rm m,0}\gamma\left(\frac{\rho_{\rm m}}{\rho_{\rm m,0}}\right)^{\frac{3\omega+4}{3\omega+3}}
\quad\textnormal{as}\quad
{\bar w}_{\rm DE}\sim\frac{1}{3\omega+3}\\
\quad\textnormal{for}\quad z_{\rm PDL} \ll z \lesssim z_{\rm eq}.
\end{aligned}
\end{equation}
This plateau is persistent all the way to the large redshift values at which either the radiation becomes dominant over pressureless matter ($z\sim z_{\rm eq}$), or the contribution from the expansion anisotropy becomes significant. The plateau of ${\bar w}_{\rm DE}$ is located within the range $0\leq{\bar w}_{\rm DE}\leq \frac{1}{3}$ depending on the value of $\omega\geq 0$. For $\omega=0$, the effective DE behaves like an extra relativistic degree of freedom as ${\bar w}_{\rm DE}=\frac{1}{3}$ during this period and hence requires special attention. On the other hand, as it is discussed above, to obtain a viable model (viz., which deviates from the standard $\Lambda$CDM model only slightly), we expect $\omega\gtrsim 10$ and this places the plateau in a value in the range $0<{\bar w}_{\rm DE}\lesssim 0.03$. During this period, the evolution of the anisotropy of the effective DE can approximately be given by \eqref{eqn:anisoEoSdust}. We note that, although $\Delta w_{\rm DE}$ increases as $(1+z)^{\frac{3}{2}}$ during this period, the EoS parameter of the effective DE would not achieve a significant anisotropy during this period even under the weakest observational constraints we give in Sect.~\ref{obsresults}.

At very/sufficiently large redshifts, viz., when the only dominant contribution is from the expansion anisotropy in \eqref{wdedef}, the ratio of the energy densities of the effective DE and the expansion anisotropy freezes out;
\begin{equation}
\begin{aligned}
\label{rhode2}
\rho_{\rm DE}\sim (\gamma-1)\rho_{\sigma^2}\quad\textnormal{and}\quad\bar w_{\rm DE}\sim\frac{3\omega+5}{3\omega+3}.
\end{aligned}
\end{equation}
We note that, except the GR limit $\omega\rightarrow\infty$ (i.e., $\gamma\rightarrow 1$), the mean/volumetric EoS parameter of the effective DE yields a second plateau placing in the range $1\lesssim \bar w_{\rm DE}\lesssim\frac{5}{3}$ for $\omega\geq0$, and the effective DE behaves like the expansion anisotropy with a value of magnitude of ratio in the range $0<|\rho_{\rm DE}/\rho_{\sigma^2}|\lesssim \frac{1}{2}$ but yields negative energy density since the coefficient $\gamma-1$ is negative for $\omega\geq0$. According to this the energy density of the effective DE changes sign and we calculate from \eqref{rhodesol} that this occurs at
\begin{equation}
\label{zpole}
z=z_{\rm DE, pole}\sim \left(\frac{\gamma}{1-\gamma}\,\, \frac{\Omega_{{\rm m},0}}{\Omega_{{\sigma^2},0}}\right)^ {\frac{\omega+1}{3\omega+4}}-1,
\end{equation}
at which its EoS parameter exhibits a pole since at that redshift its energy density becomes zero. For $z>z_{\rm DE, pole}$, $\Delta w_{\rm DE}$ asymptotically approaches to \eqref{eqn:anisoEoSaniso} with increasing redshift and the EoS of the effective DE will be significantly anisotropic, viz., we have $1<\bar w_{\rm DE}\leq\frac{5}{3}$ and  $-2\leq\Delta w_{\rm DE}<-1.2$ for $\omega\geq0$. However, for $\omega\geq0$, this interesting behavior is always irrelevant to the observable Universe since it occurs when the Universe is strongly dominated by the expansion anisotropy: Substituting $z=z_{\rm DE, pole}$ into $\frac{\rho_{\sigma^2}}{\rho_{\rm m}}=\frac{\Omega_{\sigma^2,0}}{\Omega_{\rm m,0}} (1+z)^{3+\frac{2}{1+\omega}}$, which may be read off from \eqref{hsquared}, we obtain 
\begin{equation}
\frac{\rho_{\sigma^2}(z=z_{\rm DE,pole})}{\rho_{\rm m}(z=z_{\rm DE,pole})}=\left(\frac{\gamma}{1-\gamma}\right)^{\frac{3\omega+5}{3\omega+4}}\left(\frac{\Omega_{\rm m,0}}{\Omega_{\rm \sigma^2,0}}\right)^{\frac{1}{3\omega+4}}.
\end{equation}
We note that, in a realistic cosmological setup, say, $\omega\gtrsim10$ (even for $\omega\geq0$) and $\Omega_{\rm m,0}\gg\Omega_{\sigma^2,0}$, this ratio is obviously much larger than unity, implying that $z_{\rm DE,pole}$ occurs always at a redshift at which the expansion anisotropy is already dominant.
If we consider the presence of radiation as well, then we have the following: During the radiation domination BD gravity mimics GR (as Jordan field $\varphi$ freezes out) so that the effective DE will be constant (i.e., mimics cosmological constant) while the expansion anisotropy keeps on growing as $\rho_{\sigma^2}\propto (1+z)^6$ as $z$ increases. Eventually, the expansion anisotropy will dominate over radiation and the dynamical effective DE --viz., $\rho_{\rm DE}\propto \rho_{\sigma^2}(\gamma-1)$ as may be seen from \eqref{rhodesol} by setting $\rho_{\rm M}=0=\rho_{\rm m}$-- will show up again and then its EoS parameter will respectively realize a pole and a plateau that follows that pole as $z$ increases. Thus, in a realistic setup, the pole and the second plateau features of the effective DE occur in the expansion anisotropy dominated universe that takes place before (in terms of time) the radiation dominated universe and hence they are irrelevant to the observable universe.

\subsubsection{Modified expansion anisotropy}
\label{anidiscuss}

Almost all of the model dependent observational constraints on the expansion anisotropy in the literature, e.g., \cite{Barrow:1976rda,Pontzen16,Saadeh:2016bmp,Saadeh:2016sak,Akarsu:2019pwn}, consider GR+isotropic sources leading to the steep redshift dependency of the energy density corresponding to the shear scalar as $\rho_{\sigma^2} \propto (1+z)^{6}$ and thereby give $\Omega_{\sigma^2,0}\lesssim 10^{-21}$. On the other hand, the direct/model independent observational constraints from, e.g., type Ia SNe observations are rather weak, e.g., $\Omega_{\sigma^2}\lesssim 10^{-4}$ for $z\sim 0$ \cite{Campanelli:2010zx,Wang:2017ezt}. Such large values may be possible provided that the redshift dependence of $\rho_{\sigma^2}$ is modified properly (viz., is made modest), which may be done, in principle, by either introducing an anisotropic source, e.g., anisotropic DE, in GR or considering usual cosmological sources in modified theories of gravity such as BD gravity that leads to an effective anisotropic source.

\begin{figure}[t!]
\captionsetup{justification=raggedright,singlelinecheck=false,font=footnotesize}
\par
\begin{center}
   \includegraphics[width=0.4\textwidth]{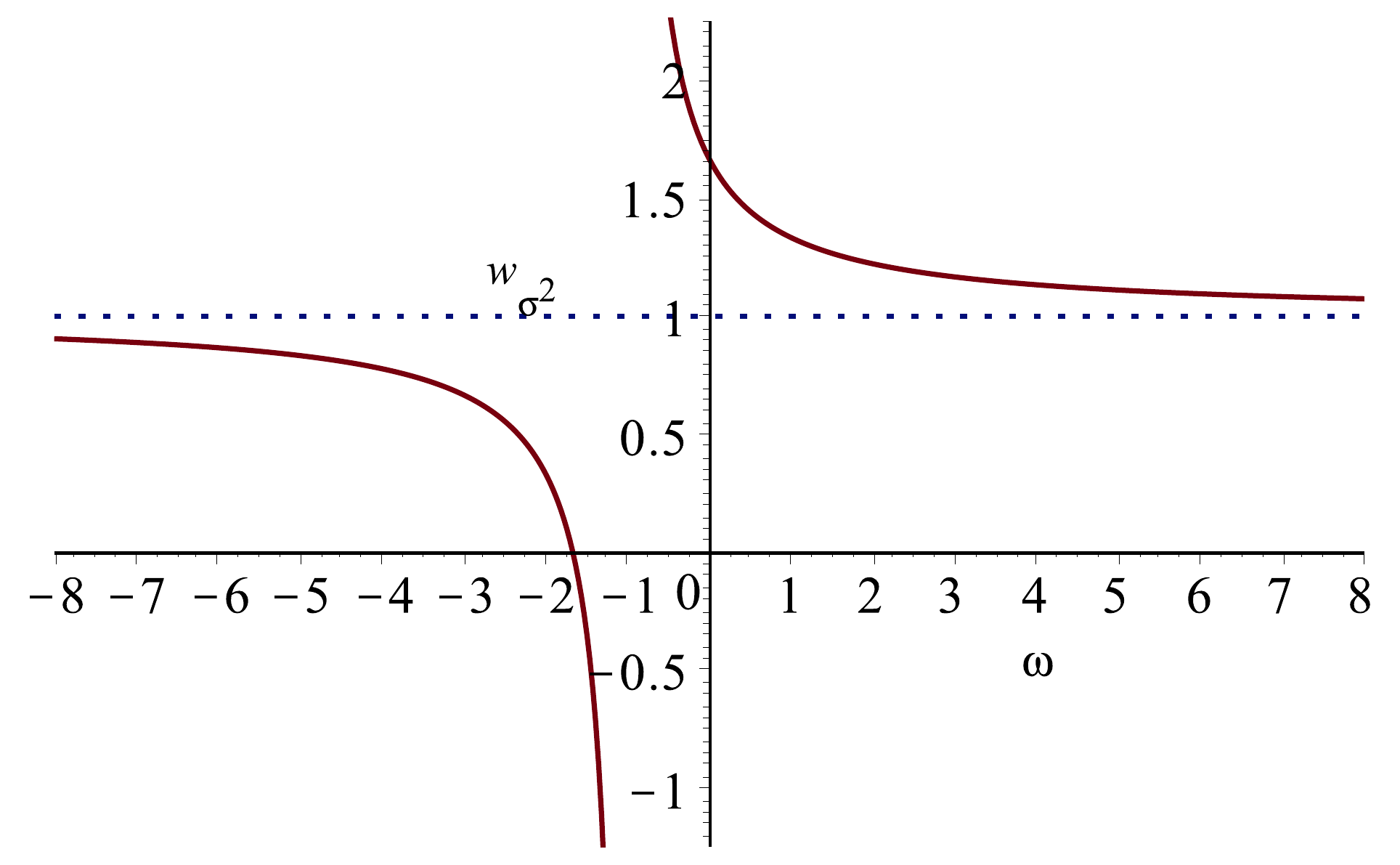}
\end{center}
\caption{Effective equation of state parameter corresponding to the expansion anisotropy $w_{\sigma^2}$ versus Brans-Dicke parameter $\omega$. $w_{\sigma^2}$ exhibits a pole at $\omega=-1$, and $w_{\sigma^2}\rightarrow1$ as $|\omega|\rightarrow\infty$.}
\label{fig:wsigma2}
\end{figure}

As we have shown in Sect.~\ref{radsol} that switching to the massive BD gravity \eqref{eq:action} with $\omega\geq0$ results in having more stringent constraints on the expansion anisotropy [see \eqref{omsigma0constraint} and discussion that follows]. These constraints in fact can be weakened for $\omega<-1$ (since the expansion anisotropy in this case yields flatter redshift dependency) and even vastly for $-\frac{5}{3}<\omega<-1$, which however would cost to large deviations from GR, hence from $\Lambda$CDM model, such that in this case we will have rapidly varying cosmological gravitational coupling strength [see \eqref{eq:Geff}] or, in the approach realized in \eqref{stdfriedwithrho}, to an effective DE significantly deviating from $\Lambda$ [see \eqref{rhodesol} along with \eqref{eq:sigmarho}]. It may be interesting to discuss a bit more on how the expansion anisotropy can be eased down in case $\omega<-1$ by considering \eqref{stdfriedwithrho} based on the approach that can be described by \eqref{einsteinian}. We see from \eqref{rhode2}-- or directly from \eqref{eq:sigma}-- that the effective EoS parameter corresponding to the expansion anisotropy, viz., shear scalar, is
\begin{equation}
w_{\sigma^2}=1+\frac{2}{3\omega+3}.
\end{equation}
We first note that it approaches Zeldovich fluid, i.e., $w_{\sigma^2}\rightarrow1$, at the GR limit $|\omega|\rightarrow\infty$ and, for $\omega\geq0$ (finite), is in the range $1< w_{\sigma^2}\leq\frac{5}{3}$, which is stiffer than the Zeldovich fluid. Zeldovich fluid yields the most rigid EoS parameter ($w=1$) compatible with the requirements of relativity theory \cite{zel61}, which in turn implies that we can not have a minimally interacting source whose energy density redshift dependence is steeper than $(1+z)^{6}$ within GR, whereas we have it in BD gravity from the expansion anisotropy as an \textit{effective source} provided that $\omega>-1$ as can be seen from \eqref{eq:sigmarho}. We depict $w_{\sigma^2}$ versus $\omega$ in Fig.~\ref{fig:wsigma2}.

For $\omega\leq-\frac{4}{3}$, the EoS parameter spanning the range $-1\leq w_{\sigma^2}<1$, which is familiar from the conventional cosmology: (i) $\omega=-2$ ($w_{\sigma^2}=\frac{1}{3}$), the expansion anisotropy mimics radiation, viz., plays the role of an extra relativistic species. (ii) $\omega=-\frac{5}{3}$ ($w_{\sigma^2}=0$), it mimics dust, viz., CDM. (iii) $\omega=-\frac{3}{2}$ --the scale invariant limit-- ($w_{\sigma^2}=-\frac{1}{3}$), it mimics constant negative spatial curvature (hyperboloid space). (iv) $\omega=-\frac{4}{3}$ ($w_{\sigma^2}=-1$), it mimics $\Lambda>0$. It might be interesting to remind here that $-\frac{3}{2}<\omega<-\frac{4}{3}$ corresponds to the range in which the universe exhibits accelerating expansion in the presence of only pressureless matter (see footnote 4). The BD parameter $\omega\sim-\frac{4}{3}$ appears in $d$-branes string models \cite{Duff:1994an,Lidsey:1999mc} and hence it is conceivable that such models would predict $\sigma^2\sim{\rm const}$. For $-\frac{4}{3}<\omega<-1$, we have $w_{\sigma^2}<-1$, i.e., the expansion anisotropy mimics phantom sources implying that the expansion anisotropy decreases/increases with increasing/decreasing redshift in contrast to the all other cases. The case $\omega=-1$ (the low energy effective string action limit of BD gravity \cite{Duff:1994an,Lidsey:1999mc}) is interesting that $w_{\sigma^2}$ exhibits a pole, namely, $\lim_{\omega \, \to \, -1^+}w_{\sigma^2}=+\infty$ and $\lim_{\omega \, \to \, -1^-}w_{\sigma^2}=-\infty$, which imply that we must set $\sigma^2=0$ in this case (at least in our solution). In relevance with this, in the case of $\omega>-1$ but $\omega\approx-1$, $w_{\sigma^2}$ will be extremely large, implying that in this case the presence of the expansion anisotropy today or at any moment in the observable Universe will require extreme fine tuning. Finally, for $\omega=0$ --the lowest value we consider in this study-- we have $\omega_{\sigma^2}=\frac{5}{3}$ and for $\omega\gtrsim 10$, which, as we discussed above, is required for small deviations from the $\Lambda$CDM model (by keeping isotropic RW metric), we have $1<w_{\sigma^2}\lesssim 1.06$. These imply that considering BD gravity with $w\gtrsim 10$ rather than GR as the law of gravity will additionally lead to small modification in the evolution of the expansion anisotropy (viz., a slightly steeper redshift dependency of the shear scalar compared to the one in GR) in the anisotropic generalization of the $\Lambda$CDM model in contrast to DE models that can be described by scalar field/s within GR.

\section{Constraints from recent cosmological data}
\label{obsresults}
 In this section we perform a parameter estimation and provide observational constraints on the free parameters of the model:  the  pressureless  matter  density  parameter  today $\Omega_{\rm m,0}$,  the baryon density parameter today $\Omega_{\rm b,0}h^2$, and the dimensionless Hubble constant $h$ as well as the two free parameters $\omega$ and $\Omega_{\sigma^2,0}$ that account for the \textit{anisotropic BD extension} of the $\Lambda$CDM model. We consider $H(z)$ given in \eqref{eq:fried1zlate} rather than \eqref{stdfriedwithrho}, since the former one contains radiation (viz., $\Omega_{\rm r,0}$) as well and can be safely used all the way to the recombination redshift (covering last scattering surface), so that we could include CMB data in our analysis. For the derived parameters we present the ones relevant to the effective DE. We consider \eqref{stdfriedwithrho}, where the effective DE is defined in accordance with the exact explicit solution in redshift given in Sect.~\ref{lcdmsol} and the approach described by \eqref{einsteinian}. In this way, even though this solution ignores the presence of radiation, we could go further and investigate the dynamics of the model under consideration in terms of redshift for $z<z_{\rm eq}$ (during which radiation is negligible) analytically in the light of the observational constraints. We keep in mind that $H(z)$ throughout this study is averaged over the volumetric expansion rate of the Universe and hence the method we use constrains the expansion anisotropy through its contribution to the average expansion rate of the Universe (viz., the shear scalar $\sigma^2$, which contributes to the $H(z)$ directly via the term $\rho_{\sigma^2,0}(1+z)^{6+\frac{2}{1+\omega}}$ and also indirectly via its effect on the $\rho_{\rm DE}$ as a result of BD theory).
 
 In order to perform the parameter space exploration, we make use of a modified version of the simple and fast Markov Chain Monte Carlo (MCMC) code that computes expansion rates and distances from the Friedmann equation, named SimpleMC \cite{Anze, Aubourg:2014yra}. 
For an extended review of cosmological
parameter inference see \cite{EPadilla}. The code uses a compressed version of the Planck data (PLK), where the CMB is treated as a ``BAO experiment" at redshift $z=1090$, measuring the angular scale of the sound horizon at that time, a recent analysis of Type Ia supernova (SN) data dubbed Joint Light-curve Analysis (JLA) compressed into a piece-wise linear
 function fit over 30 bins spaced evenly in $\log z$, and high-precision Baryon Acoustic Oscillation measurements (BAO), from the comoving angular diameter distance only (viz., growth rate measurement is not considered), the Hubble distance and the volume averaged distance, at different redshifts up to $z=2.36$. For a more detailed description about the datasets used see \cite{Aubourg:2014yra}. We also include, independent of/separate from other data sets, a collection of currently available data on $H(z)$ obtained from cosmic chronometers ($H$) (see \cite{Gomez-Valent:2018hwc} and references therein). Table~\ref{tab:priors} displays the parameters used throughout this paper along with the corresponding flat priors; derived parameters labeled with $^*$.
 
 \begin{table}[t!]
\captionsetup{justification=raggedright,singlelinecheck=false,font=footnotesize}
\footnotesize
\caption{Constraints on the anisotropic Brans-Dicke extension of the standard $\Lambda$CDM model using the combined data sets PLK+BAO+SN+$H$. For two-tailed distributions, the results are given in $1\sigma$ and for one-tailed  distributions, the results are given in $2\sigma$. $\Lambda$CDM column corresponds to GR limit (Brans-Dicke gravity with $\omega\rightarrow \infty$) and the shear scalar $\sigma^2=0$ (isotropic). Parameters and ranges of the uniform priors assumed in our analysis and derived parameters are labeled with~$^*$.}
\begin{tabular}{cccc}
\cline{1-4}\noalign{\smallskip}
\vspace{0.05cm}
Parameter   &    Anisotropic BD &  $\Lambda$CDM   &  Priors   \\
 \hline
\vspace{0.05cm}
$\Omega_{\rm m,0}$			& $0.3002(67) $ & $0.3021(64) $	& $[0.05,1.5]$ 	\\
\vspace{0.05cm}
$\Omega_{\rm b,0}h^2$			& $0.02241(16)$ & $0.02244(15)$	 &[0.02, 0.025]\\
\vspace{0.05cm}
$h$	& $0.6848(65)$ &	 $0.6817(49)$   	&   $[0.4, 1.0]$ 	\\
\vspace{0.05cm} 
$\log_{10}\omega$			& $>1.69$			&	 [$\omega\rightarrow+\infty$] 	& [0, 6]\\
\vspace{0.05cm}
$\log_{10}\Omega_{\sigma^2,0}$	& $< -8.48$		&	 [0]		& $[-12, 0]$\\
\hline
\vspace{0.05cm}
$^*{\bar w}_{\rm DE,0}$			& $>-1.0044$		&	 $[-1]$ 	&	\\
\vspace{0.05cm}
$^*\Delta w_{\rm DE,0}$		& $<4.23\times 10^{-7}$&	 $[0]$ 	&	\\

\vspace{0.05cm}
$^*z_{\rm PDL}$		& $0.64839(58)$& -- 	&\\

\vspace{0.05cm}
$^*\log_{10}z_{\rm DE,pole}$		& $4.80(58)$ & --  				&\\

\vspace{0.05cm}
$^*z_{\rm eq}$		& $3368.62\pm 30.89$ & $3359.45\pm 24.14$ 	 	&\\

\vspace{0.05cm}
$^*2\omega M{^2}$ [$10^{-66}$eV$^2$]		& $4.48\pm 1.22$ &	($\Lambda$) $4.48 \pm 1.25$  				&\\

\vspace{0.05cm}
$^*M$ [$10^{-34}$eV]	 	&		$<1.51$			&--		&\\

\hline
\vspace{0.05cm}
$-2\Delta\ln \mathcal{L}$				&  $-0.62$ 	&	 0				& --  \\
\hline
\end{tabular}
\label{tab:priors}
\end{table}
 
In the analysis, we assume radiation content by including three neutrino species ($N_{\rm eff}=3.046$) with minimum allowed mass $\sum m_{\nu}=0.06\, {\rm eV}$ theoretically well determined within the standard model of particle physics and the density parameter of radiation $\Omega_{{\rm r},0}=2.469\times 10^{-5} h^{-2}(1+0.2271 N_{\rm eff})$, where dimensionless Hubble constant $h=H_0/100\, {\rm km\,s}^{-1}\,{\rm Mpc}^{-1}$ \cite{Komatsu:2010fb}. Throughout the analysis we assume flat priors over our sampling parameters: $\Omega_{{\rm m},0}=[0.05,1.5]$, $\Omega_{{\rm b},0} h^2=[0.02,0.025]$ and $h=[0.4,1.0]$. Whereas priors for the two parameters that characterize the anisotropic BD extension of the standard $\Lambda$CDM model are given by $\log_{10}\omega =[0,6]$ and $\log_{10}\Omega_{\sigma^2,0}= [-12,0]$. The photon energy density today $\rho_{{\gamma},0}$ is not subject to our analysis since it is well constrained, such that it has a simple $\rho_{\gamma}=\frac{\pi^2}{15} T_{\rm CMB}^4$ relation with the CMB monopole temperature  \cite{Dodelson03}, which is very precisely measured to be $T_{{\rm CMB},0}=2.7255\pm 0.0006\,{\rm K}$ \cite{Fixsen09}. Table~\ref{tab:priors} summarizes the observational constraints on the free parameters as well as the derived parameters of the model under consideration using the combined datasets PLK+BAO+SN+$H$; and for comparison those parameters used on the standard $\Lambda$CDM model.

\begin{figure*}[t!]
\captionsetup{justification=raggedright,singlelinecheck=false,font=footnotesize}
\par
\begin{center}
\includegraphics[trim = 1mm  1mm 1mm 1mm, clip, width=6.cm, height=4.5cm]{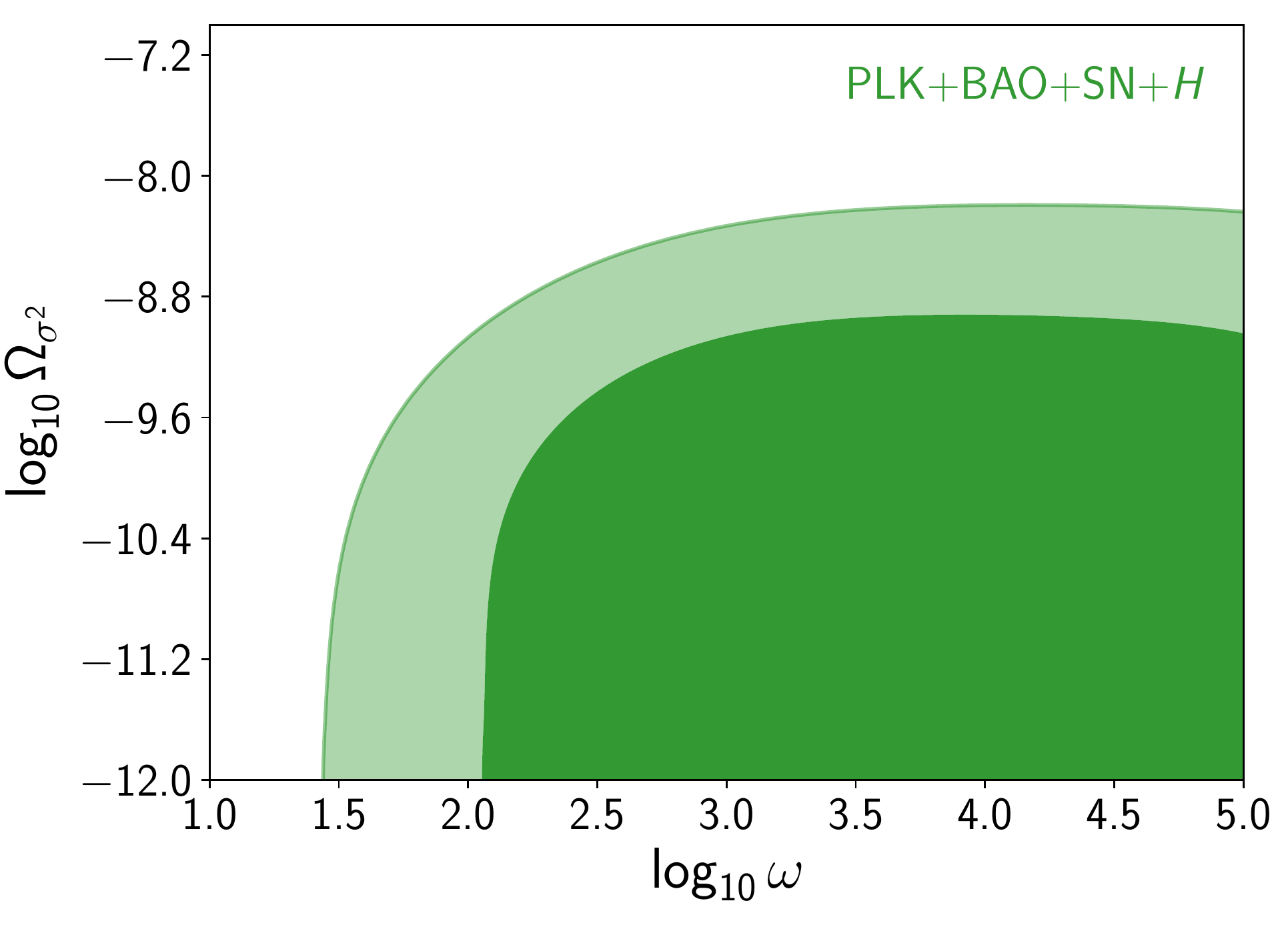}
  \includegraphics[trim = 1mm  1mm 1mm 1mm, clip, width=6.cm, height=4.5cm]{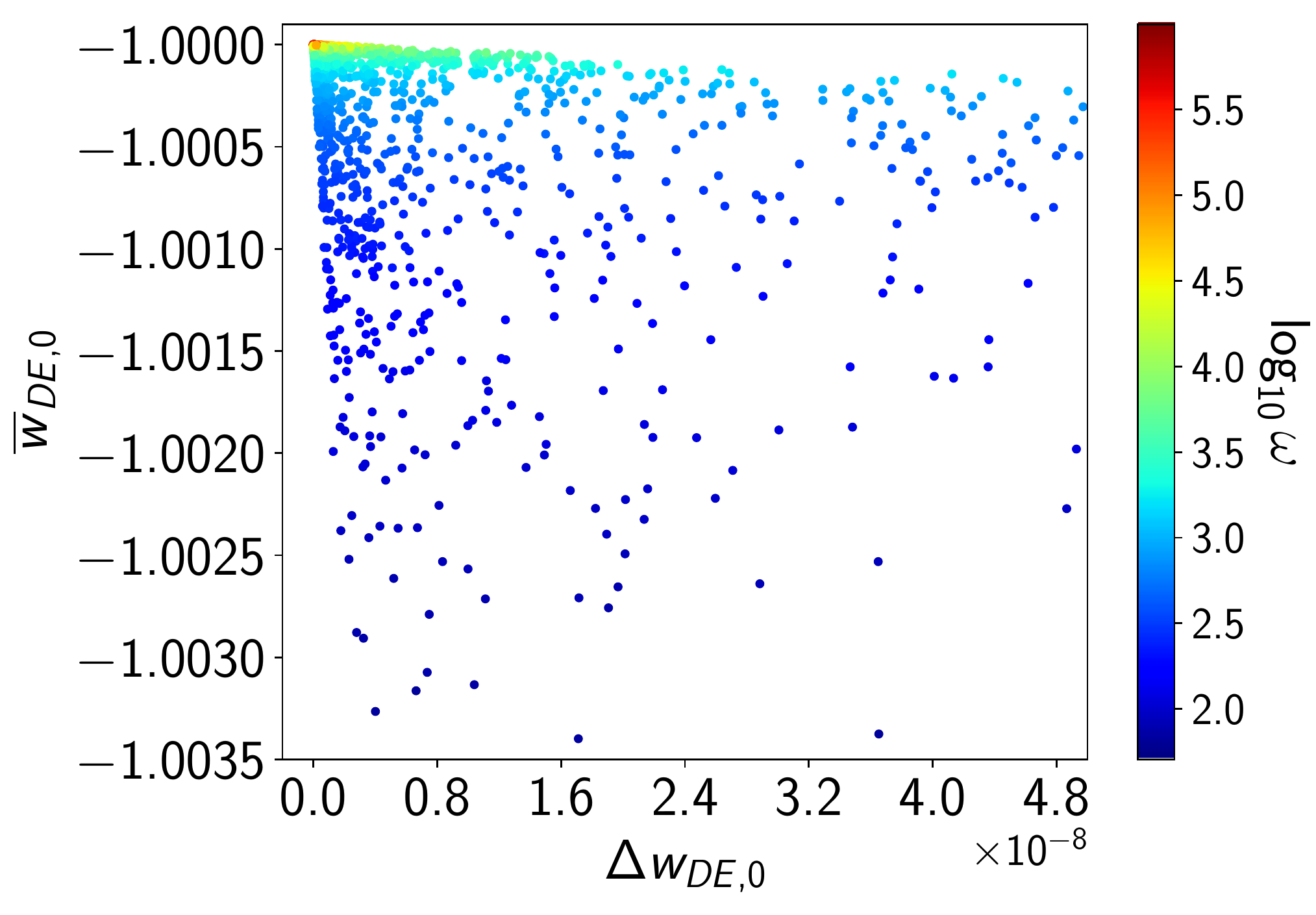}
\end{center}
\caption{ 
\textbf{(Left panel)}: 2D marginalized posterior distribution of the parameters $\log_{10}\Omega_{\sigma^2}$ and $\log_{10}\omega$ that determine the extension from the standard $\Lambda$CDM model.
The 2D constraints are plotted with 1$\sigma$ and 2$\sigma$  confidence contours. 
\textbf{(Right panel)}: the 3D posterior distribution in the
$\{\bar{w}_{\rm DE,0}, ~\Delta w_{\rm DE,0}, ~\log_{10}\omega\}$ subspace, where the colour code indicates the value of $\log_{10}\omega$ 
using the colour bar.}
\label{aniso}
\end{figure*}

\begin{figure*}[t!]
\captionsetup{justification=raggedright,singlelinecheck=false,font=footnotesize}
\par
\begin{center}
  \includegraphics[trim = 1mm  1mm 1mm 1mm, clip, width=6.4cm, height=4.5cm]{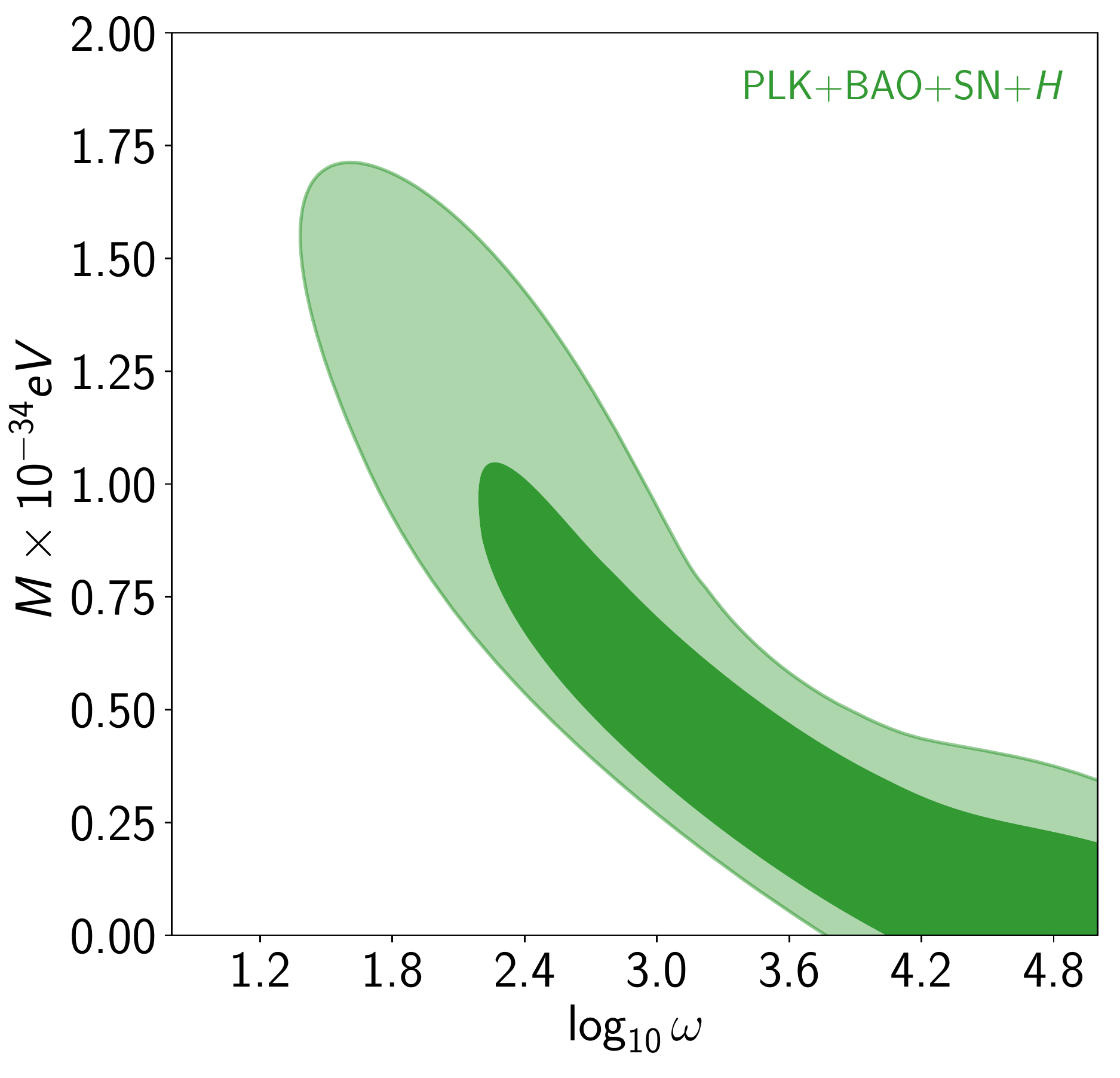}
   \includegraphics[trim = 1mm  1mm 1mm 1mm, clip, width=6.cm, height=4.5cm]{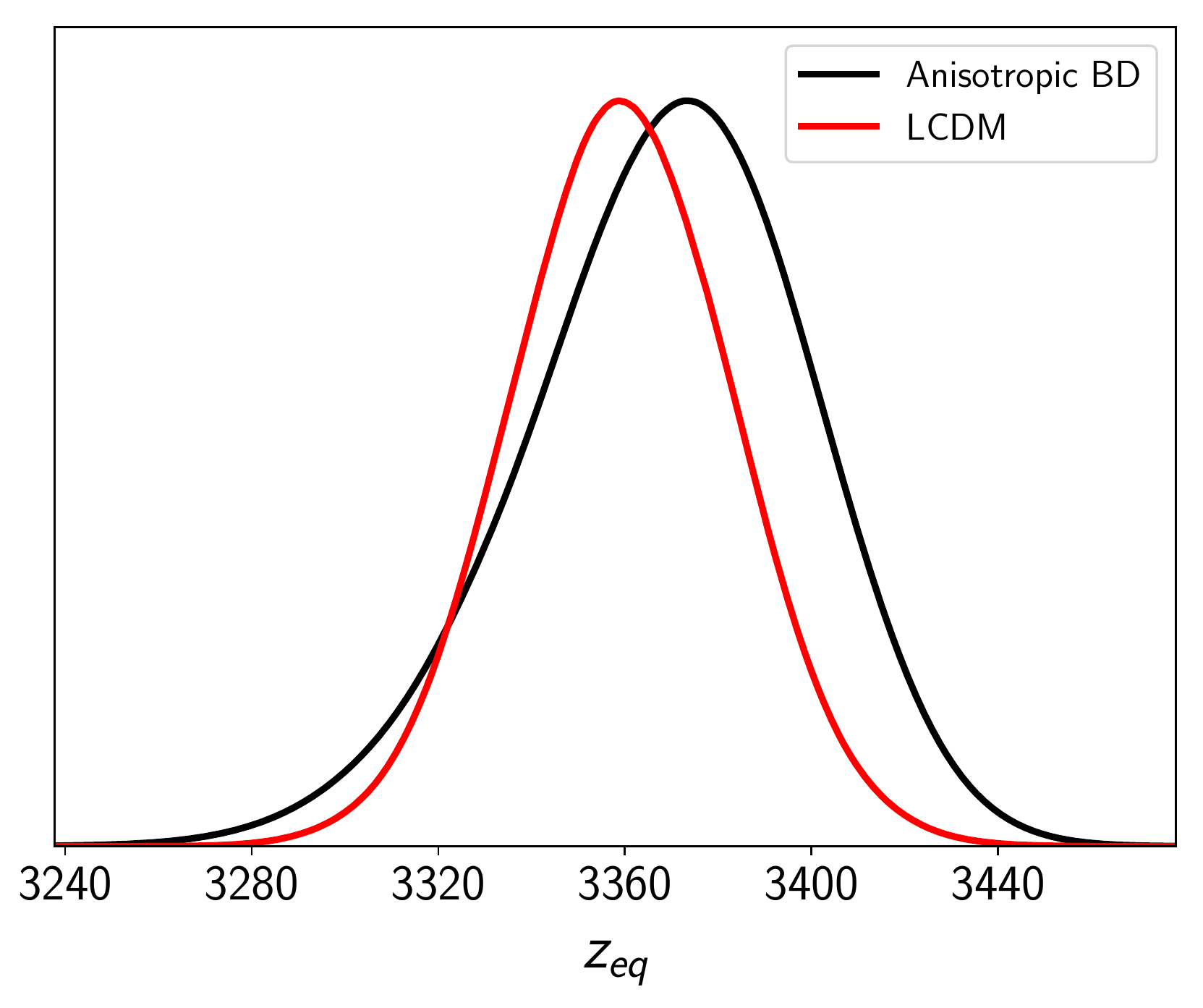}

\end{center}
\caption{ 
\textbf{(Left panel)}: 2D marginalised posterior distribution of the parameters corresponding to the mass of the Jordan field $M$ and the Brans-Dicke parameter $\log_{10}\omega$. \textbf{(Right panel)}: 1D constraints corresponding to the redshift of matter-radiation equality ($z_{\rm eq}$).
}
\label{fig:MH}
\end{figure*}

\begin{figure*}[t!]
\captionsetup{justification=raggedright,singlelinecheck=false,font=footnotesize}
\par
\begin{center}
\includegraphics[trim = 0mm  0mm 1mm 1mm, clip, width=6.5cm, height=4.2cm]{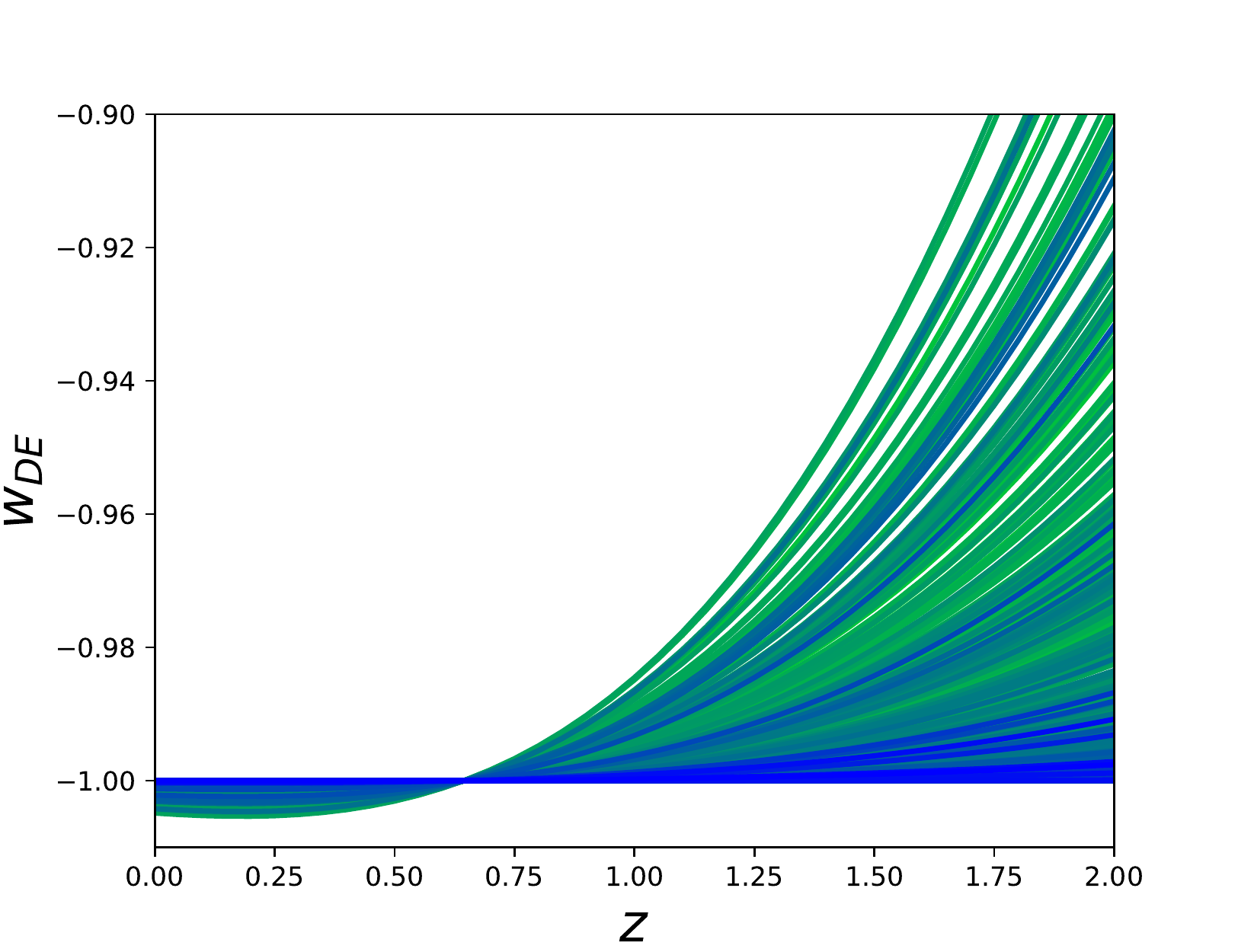}
\includegraphics[trim = 0mm  0mm 1mm 1mm, clip, width=6.5cm, height=4.cm]{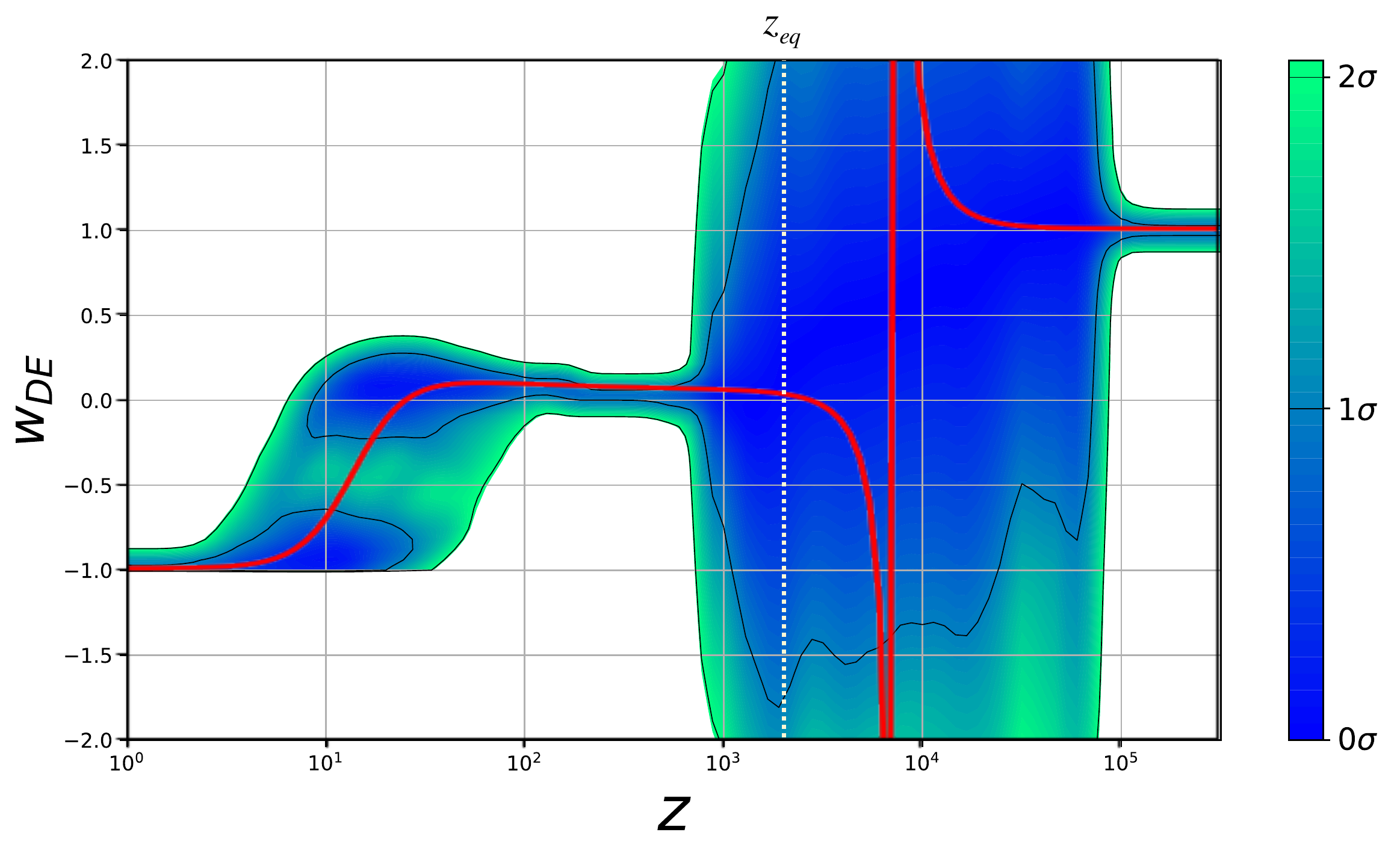}\\
\includegraphics[trim = 0mm  0mm 1mm 1mm, clip, width=6.5cm, height=4.2cm]{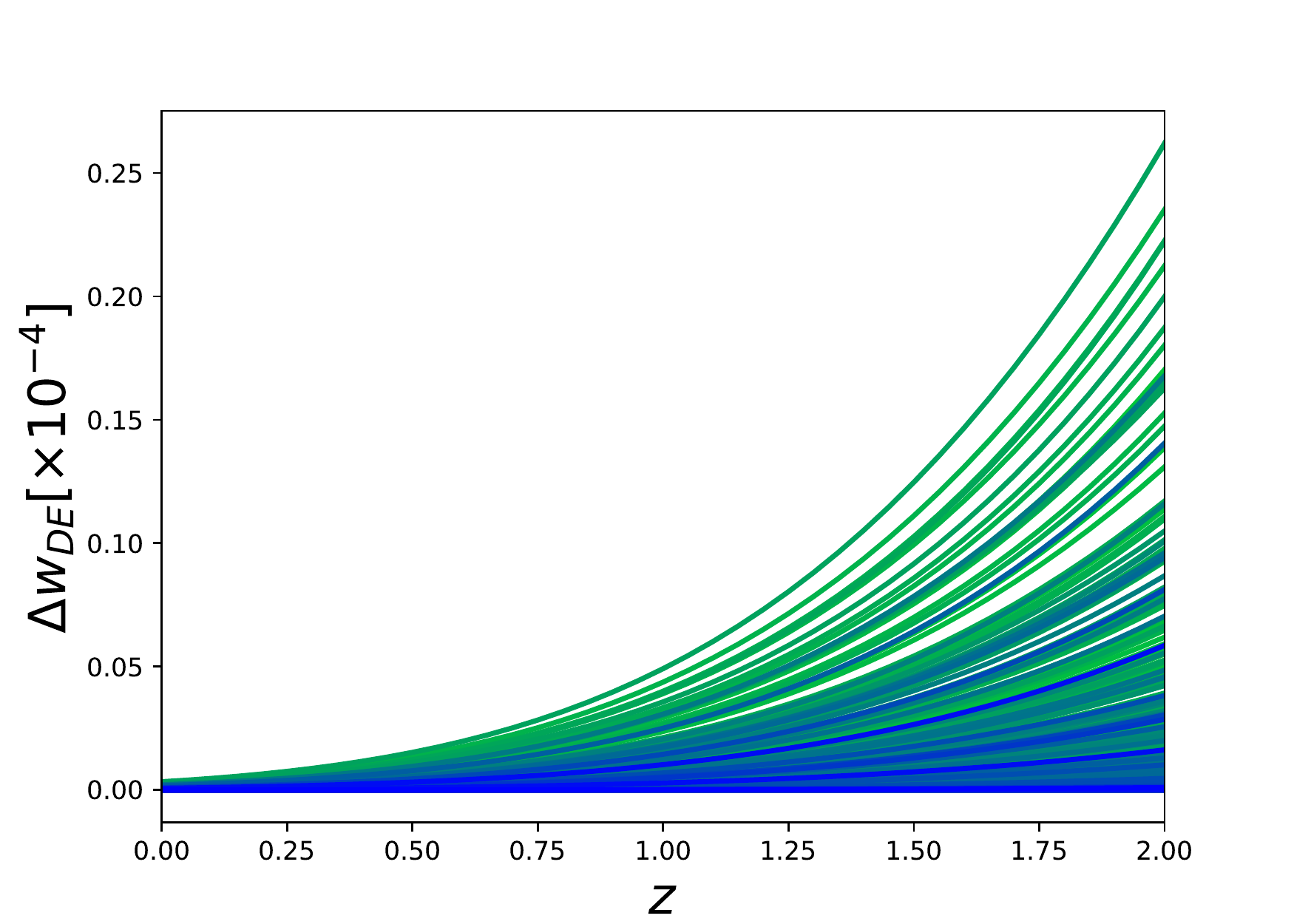}
\includegraphics[trim = 0mm  0mm 1mm 1mm, clip, width=6.5cm, height=4.cm]{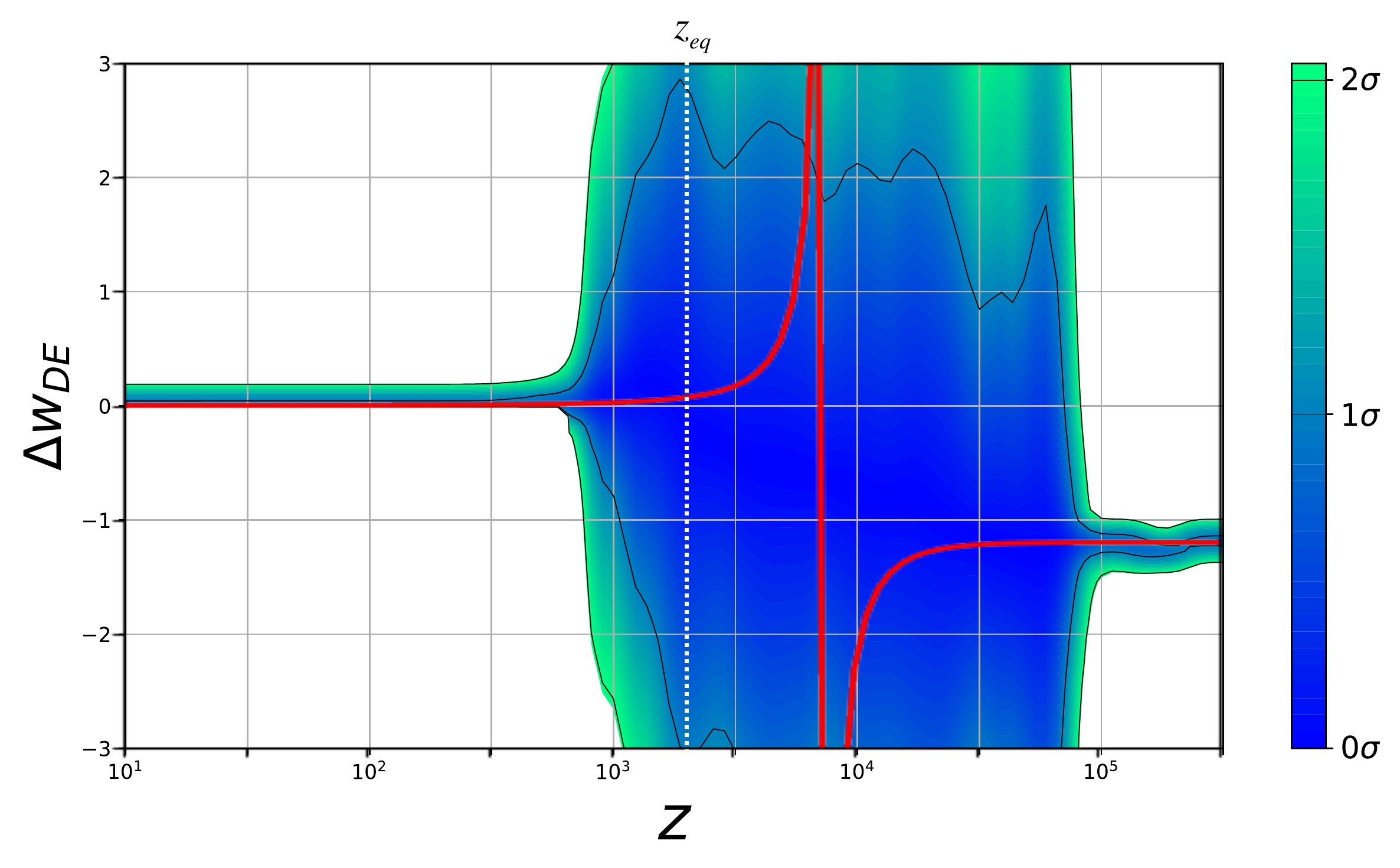}
\includegraphics[trim = 0mm  0mm 1mm 1mm, clip, width=6.5cm, height=4.2cm]{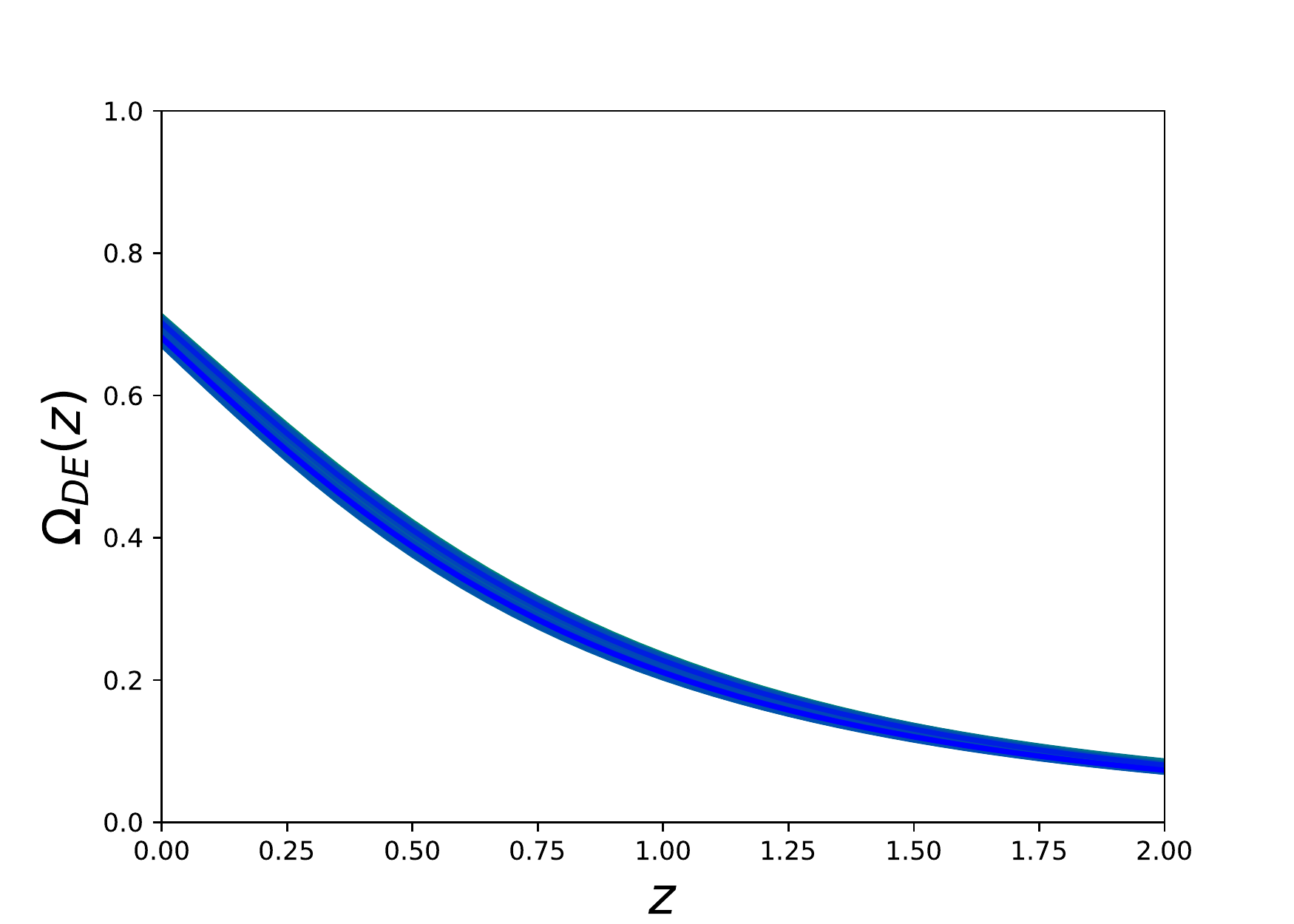}
\includegraphics[trim = 0mm  0mm 1mm 1mm, clip, width=6.5cm, height=4.cm]{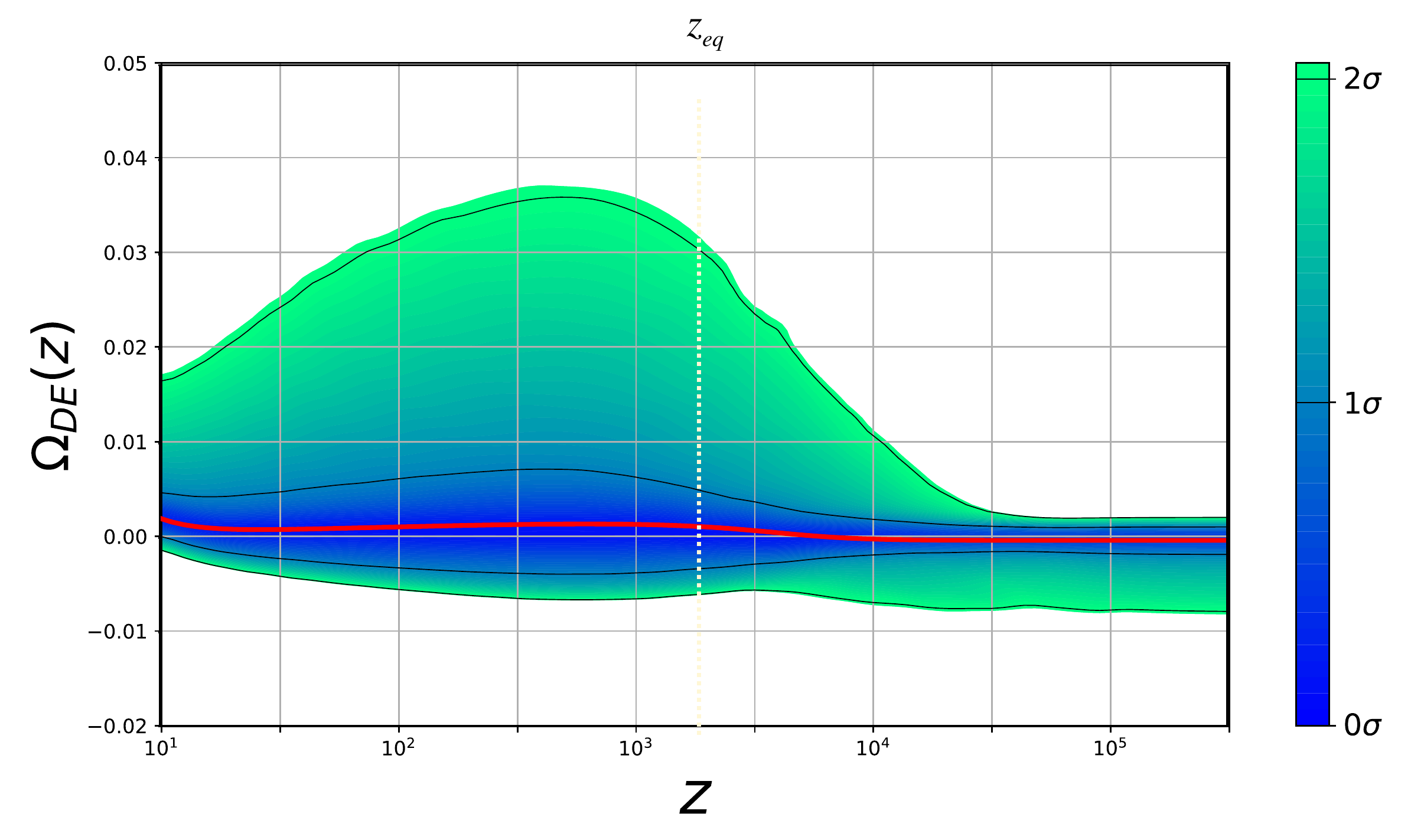}
\end{center}
\caption{The constraints on $g(z)$ as a result of the data. These show the posterior probability Pr$(g|z)$: the probability of $g$ as normalized in each slice of constant $z$, with color scale in confidence interval values. The 1$\sigma$ and 2$\sigma$ confidence intervals are plotted as black lines and red lines correspond to the best-fit values found over the analysis, while dotted lines to the redshift of matter-radiation equality. \textbf{(Top)}: $\bar w_{\rm DE}(z)$ at low $z$ values (left) and at high redshift values in log-scale (right). \textbf{(Middle)}: $\Delta w_{\rm DE}(z)$ at low $z$ values (left) and at high redshift values in log-scale (right). \textbf{(Bottom)}: $\Omega_{\rm DE}(z)$ at low $z$ values (left) and at high redshift values in log-scale (right).
}
\label{fig:wde,deltawde,omfrac}
\end{figure*}

Fig.~\ref{aniso} summarizes the constraints on the parameters that characterize the anisotropic BD extension to the standard $\Lambda$CDM model. Left panel of this figure displays 2D marginalized posterior distributions on the BD parameter $\omega$ (deviation from GR) and the density parameter corresponding to the expansion anisotropy today $\Omega_{\sigma^2,0}$ (deviation from isotropic expansion). We see from Table~\ref{tab:priors} that current observations prefer small contributions from the expansion anisotropy $\log_{10}\Omega_{\sigma^2,0} < -8.48$ (95\% C.L.) with a large BD parameter (small deviation from GR) $\log_{10}\omega >1.69$ (95\% C.L.), signaling that the anisotropic BD extension to the standard $\Lambda$CDM model should be studied indeed as a \textit{small correction} in line with what we have been claiming and basically doing so far. Right panel of Fig.~\ref{aniso} displays 3D posterior distributions in the
$\{\bar{w}_{\rm DE,0}, \Delta w_{\rm DE,0}$, $\log_{10}\omega\}$ subspace:
  the average/volumetric EoS parameter of the effective DE $\bar w_{\rm DE,0}$ and the measure of anisotropy of its EoS parameter $\Delta w_{\rm DE,0}$ coloured code with $\log_{10}\omega$. The former two are derived parameters that determine deviations of the effective DE from $\Lambda$ ($\bar w_{\rm DE}=-1$ and $\Delta w_{\rm DE}=0$) today and are controlled by $\omega$. We notice that for large values $\omega$ 
 (color coded with red) 
 we recover the $\Lambda$CDM case at the present time, where $\Delta w_{\rm DE,0}=w_{{\rm DE},y,0}-w_{{\rm DE},x,0} \to 0$ and $\bar w_{\rm DE,0} \to -1$,  as expected by our previous analysis. We would like to remind here that in our observational analysis only positive values of $\omega$ were allowed. The left panel of Fig.~\ref{fig:MH} shows that current data impose upper bounds on the mass of the Jordan field given by $M<1.51 \times 10^{-34}$eV (95\% C.L.). The correlation observed on the parameters $\omega$ and $M$ is an interesting point to note. Even though both parameters have only one tail constraints, the effective cosmological constant $2\omega M^2$ --a component of the effective DE, see \eqref{rhoss} and \eqref{rhodesol}-- is well limited by $2\omega M{^2}=4.48\pm 1.22\times 10^{-66}$eV$^2$ and in agreement with the current value of $\Lambda=4.48 \pm 1.25\times 10^{-66}$eV$^2$, see Table~\ref{tab:priors}. The extra two parameters from the anisotropic BD extension to the $\Lambda$CDM model may shift and relax some of the constraints, as it is the case of 
$\Omega_{\rm m,0}=0.3002 \pm 0.0067$, and therefore as a consequence the earlier redshift at matter-radiation equality $z_{\rm eq}=3368.62\pm 30.89$, compared to the $\Lambda$CDM model with $z_{\rm eq} = 3359.45\pm 24.14$ (see right panel of Fig.~\ref{fig:MH}).  

We depict, in Fig.~\ref{fig:wde,deltawde,omfrac}, the behaviour of the effective DE (top and middle panels) as well as the evolution of its density parameter $\Omega_{\rm DE}=\rho_{\rm DE}/3H^2$ (bottom panel) in redshift, according to the constraints obtained from observational analysis (see Sect.~\ref{sub1preliminary} for the features of the effective DE). Left panel of Fig.~\ref{fig:wde,deltawde,omfrac} displays the behaviors of $\bar{w}_{\rm DE}$, $\Delta w_{\rm DE}$ and $\Omega_{\rm DE}$ for redshifts up to $z=2$. This panel was drawn by taking random samples from the posterior distribution of the parameter-space and colour coded by its likelihood (bluer colors represent regions of higher probability). Right panel of Fig.~\ref{fig:wde,deltawde,omfrac} displays the extended behaviors of the same functions; from redshift $z=1$ to $z=10^6$ for $\bar{w}_{\rm DE}$ and from redshift $z=10$ to $z=10^6$ for $\Delta w_{\rm DE}$ and $\Omega_{\rm DE}$. Though, note that the extensions to the redshift values $z\sim z_{\rm eq}$ and larger should be seen with a caution that these are obtained by making use of the effective DE derived in Sect.~\ref{sub1preliminary} without considering the presence of radiation. In this panel we depict the probability of a function normalized in each slice of constant $z$, with colour scale in confidence interval values. The 1$\sigma$ and 2$\sigma$ confidence intervals are plotted as black lines and red lines correspond to the best-fit values found over the analysis. At low redshifts, the upper-left panel shows that the crossing of the PDL occurs at $z_{\rm PDL}=0.64839\pm0.00058$, just as we concluded from the discussion that follows equation \eqref{rhomsignz}. For redshift values $z<z_{\rm PDL}$, the effective DE exhibits phantom behaviour $\bar{w}_{\rm DE}<-1$ as may be seen from \eqref{wde0} or \eqref{rhode1} and for higher redshift values, $z>z_{\rm PDL}$, $\bar{w}_{\rm DE}>-1$ as may be seen from \eqref{wdedefmod} and \eqref{rhode2}. Evolving DE with an EoS parameter being below $-1$ at present, evolved from $w>-1$ in the past is named as quintom DE. We obtain the quintom DE as the upper-left panel shows, whereas, the explicit construction of quintom scenario is more difficult than other dynamical DE models, due to a no-go theorem  which forbids the EoS parameter of a single perfect fluid or a single scalar field to cross the $w=-1$ boundary \cite{Cai:2009zp}. This property is distinctive that single-scalar-field models with canonical kinetic term are not allowed to satisfy, which also lead to that the Hamiltonian is unbounded from below. Interestingly this property is also achieved throughout some model independent analyses \cite{Vazquez:2012ce,Hee:2016ce,Zhao:2017cud}. The middle left panel shows that, as can be deduced from \eqref{deltaw1now} and \eqref{eqn:anisoEoSdust}, the effective DE becomes slightly more anisotropic with the increasing redshift, viz., the anisotropy of the EoS parameter $\Delta w_{\rm DE}$ increases only about an order of magnitude from its current value $\Delta w_{\rm DE,0}\lesssim 10^{-7}$ (see Table~\ref{tab:priors}) to the one at $z=2$. Looking at the right panel we see that after few redshifts, as pressureless matter becomes dominant (see in the bottom panel that $\Omega_{\rm DE}\sim 0.1$ for $z\sim 1.75$ and $\Omega_{\rm DE}\sim 0$ for $z\gtrsim 10$), $\bar{w}_{\rm DE}$ starts to noticeably climb up and settles in the first plateau of $\bar{w}_{\rm DE}\sim 0$ (see \eqref{wdedefmod} and paragraph covering it) lying between $z\sim 50$ and $z\sim z_{\rm eq}$ (viz., throughout the pressureless matter dominated epoch). There is a period in this plateau during which $\bar w_{\rm DE}>0$ [see \eqref{wdedefmod}], viz., $\Omega_{\rm DE}$ increases (i.e., the energy density of the effective DE increases faster than of the pressureless matter) with  increasing redshift. However, as can be seen in the bottom right panel, $\Omega_{\rm DE}$ during this period can never grow up to considerable values, viz., remains less than a percent at 1$\sigma$ C.L. and few percents at 2$\sigma$ C.L.. The anisotropy of the effective DE keeps on increasing with increasing redshift approximately in accordance with \eqref{eqn:anisoEoSdust} during this plateau, but it remains positive definite and insignificant (e.g., $\Delta w_{\rm DE}\lesssim 0.07$ at photon decoupling redshift $z\sim1100$) until the effective DE starts to leave this plateau as the expansion anisotropy starts to become dominant. We see that $\bar{w}_{\rm DE}$ exhibits a pole at $\log_{10}z_{\rm DE,pole}= 4.80\pm 0.58$ then settles in a new plateau of $\bar{w}_{\rm DE}\sim 1$ starting at $z\sim 10^5$ during which $\Delta w_{\rm DE}$ exhibits a similar behavior and settles down at $\Delta w_{\rm DE}\sim-1.2$, i.e., the effective DE eventually becomes highly anisotropic. However, note that this last stage of the effective DE starting just after $z\sim z_{\rm eq}$ is basically an artifact of omitting radiation in Sect.~\ref{sub1preliminary} to have explicit expression of $\rho_{\rm DE}(z)$, strickly speaking, is in fact unlikely to be realized at an observationally relevant past of the Universe (see the discussion starting with equation \eqref{rhode2} in Sect.~\ref{sub1preliminary}). 

 Instead, in a realistic setup, the radiation domination should start at $z=z_{\rm eq}$ and be maintained all the way to the redshift values at which BBN took place $z\sim z_{\rm BBN}$. During which the Jordan field is constant and hence, except an altered cosmological gravitational coupling strength, the Universe evolves exactly the same as in GR (see Sect.~\ref{radsol}), such that the effective DE mimics $\Lambda$ (viz., $\bar{w}_{\rm DE}= -1$ and $\Delta w_{\rm DE}=0 $ implying that it is isotropic and maintains its energy density value at $z\sim z_{\rm eq}$ for $z>z_{\rm eq}$) and expansion anisotropy increases as $\propto(1+z)^6$. However, although the data constrain $z_{\rm DE,pole}$ to be about 1-2 orders of magnitude larger than $z_{\rm eq}$, we see in Fig.~\ref{fig:wde,deltawde,omfrac} that, even within the 68\% C.L. error region, the abrupt behaviour of the effective DE can start at redshift values less than $z_{\rm eq}$, which implies that the expansion anisotropy in this case is too large that it will never allow radiation to be dominant in the Universe, though over pressureless matter. Moreover, even there is a region where abrupt behaviour of the effective DE starts at redshift values larger than $z_{\rm eq}$, using the constraint on $\Omega_{\sigma^2,0}$ from Table~\ref{tab:priors} we calculate that the expansion anisotropy dominates over radiation at redshift values smaller than $z_{\rm BBN}$, i.e., it will spoil the BBN processes. All these imply that the observational constraint method we performed here is able to put stringent enough constraints on neither $\Omega_{\sigma^2,0}$ nor $\omega$.  These two parameters are not independent and hence we should further investigate the upper boundary on $\Omega_{\sigma^2,0}$ and lower boundary on $\omega$, of course, particularly, by using the data providing information about the Universe at $z\gtrsim z_{\rm eq}$, such as the peak of the matter-power spectrum relevant to $z=z_{\rm eq}$ and BBN relevant to $z\sim 10^8$.
 
 \section{Some further observational consequences and discussions}
\label{sec:varGandBBN}

\subsection{Variation of cosmological gravitational coupling strength}
\label{sec:varG}

The normalized rate of change of the cosmological gravitational coupling strength in our exact solution (neglecting radiation), from \eqref{eq:Geff}, reads
\begin{equation}
\label{eqn:varg}
\frac{\dot{G}}{G}=-\frac{H}{1+\omega},
\end{equation}
which is negative definite considering $\omega\geq0$ in this study, provided that the Universe is expanding ($H>0$). The constraints we found on $\omega$ and $H_0$ (see Table~\ref{tab:priors}) predict
\begin{equation}
\label{eqn:vargval}
\bigg|\frac{\dot{G}}{G}\bigg| <1.092\times 10^{-12} \, {\rm yr}^{-1}\,\, (68\% \,{\rm C.L.}) \quad{\rm for}\quad z=0,
\end{equation}
similar to those given for $z\sim 0$ from various physical systems in which gravity is not negligible, such as the motion of the bodies of the Solar System, astrophysical and cosmological systems (see \cite{Uzan:2010pm} for a comprehensive review). Assuming this relation \eqref{eqn:varg} holds all the way to matter-radiation equality, we obtain $|\dot G/G|\lesssim 10^{-8} \, {\rm yr}^{-1}$ for $z\sim z_{\rm eq}$. On the other hand --given that $\varphi={\rm const.}$ (and hence $G={\rm const.}$) is an attractor solution when radiation is dominant-- the redshift dependence of $G$ becomes flatter w.r.t. $G\propto (1+z)^{\frac{1}{1+\omega}}$ [see  \eqref{eq:Geff}] as the radiation becomes more significant w.r.t. pressureless matter and thereby $|\dot G/G|\lesssim 10^{-8} \, {\rm yr}^{-1}$ for $z\sim z_{\rm eq}$ will never be achieved, but instead $G$ will become almost constant for $z\sim z_{\rm eq}$ and remain so for $z>z_{\rm eq}$ as long as radiation continues to be dominant. Yet, in accordance with our detailed discussion in Sect.~\ref{radsol}, \eqref{eqn:varg} is mostly valid all the way to $z_{\rm eq}$ and leads to $G_1\sim G_0(1+z_{\rm eq})^{\frac{1}{1+\omega}}$ [see \eqref{G1G0rel}]. Accordingly, we find that the constraints from the data predict the upper bound on the relative change in the strength of the cosmological gravitational coupling at $z\sim z_{\rm eq}$ w.r.t. its present time value as $\frac{\Delta G}{G_0}|_{z\sim z_{\rm eq}}=0.138\,(0.175)$ 68\% C.L. (95\% C.L.). This, in turn, implies that the strength of the cosmological gravitational coupling during the radiation dominance will be a constant $G_{1}$ satisfying the following constraints:
\begin{equation}
\begin{aligned}
\label{G1G0relnum}
G_0\leq G_{1}\lesssim 1.138\, G_0\,\,(1.175\, G_0)&  \quad 68\% \,{\rm C.L.}\, (95\% \,{\rm C.L.})
\end{aligned}
\end{equation}
for $z\gtrsim z_{\rm eq}$. These constraints, of course, are valid also for the epoch of primordial nucleosynthesis that takes place when $z_{\rm BBN}\sim 3\times 10^8$, provided that the expansion anisotropy is still insignificant at that redshift. We finally note that these constraints are similar to those obtained from CMB and BBN (see Ref. \cite{Uzan:2010pm} for a review on CMB and BBN constraints on $\Delta G/G_0$).
\subsection{Matter-radiation transition}
\label{MRequality}
We find using \eqref{eqn:eqsigmar} that, along with the constraints on the BD parameter and the radiation density parameter, the upper bound on the expansion anisotropy today, $\Omega_{\sigma^2,0}=10^{-8.48}$ (see Table~\ref{tab:priors}), leads to a lower bound on the expansion anisotropy-radiation equality ($\Omega_{\sigma^2}=\Omega_{\rm r}$) 
redshift as $z_{\rm eq,\sigma^2,r}\sim 10^3$, which is obviously not acceptable in a viable cosmological model since it implies that the expansion anisotropy dominates the Universe even at redshift values less than the matter-radiation equality ($\Omega_{\rm m}=\Omega_{\rm r} $) redshift $z_{\rm eq }\sim 3369$ (see Table~\ref{tab:priors}). Indeed, using the results given in Table~\ref{tab:priors}, we see that the Universe could be dominated by the expansion anisotropy as $\frac{\Omega_{\sigma^2,\rm eq}}{\Omega_{\rm r,eq}+\Omega_{\rm m,eq}}=\frac{\Omega_{\sigma^2,\rm eq}}{2\Omega_{\rm m,eq}}\lesssim 10^2$ at the matter-radiation equality, and thereby conclude that the upper bound on $\Omega_{\sigma^2,0}$ given above should be further reduced to values guaranteeing that the Universe is radiation and matter dominated at the matter-radiation equality.
Additionally, transition from radiation to matter domination is one of the most important epochs in the history of the Universe. This transition alters the growth rate of density perturbations: during the radiation era perturbations well inside the horizon are nearly frozen but once matter domination commences, perturbations on all length scales are able to grow by gravitational instability and therefore it sets the maximum of the matter power spectrum in GR as well as BD \cite{Liddle:1998ij}. Namely, it determines the wavenumber, $k_{\rm eq}$, of a mode that enters the horizon, $H_{\rm eq}a_{\rm eq}$, at the matter-radiation transition \cite{Liddle:1998ij,chen99}. 

In our model, $k_{\rm eq}=H_{\rm eq}a_{\rm eq}=\frac{H_{\rm eq}}{(1+z_{\rm eq})}$ can be estimated analytically by assuming $H(z)$ given in \eqref{eq:fried1zlate} holds all the way to matter-radiation equality. At this point, both the radiation and matter contribute equally to the total energy density. This gives us opportunity to reduce the constraints on $\Omega_{\sigma^2,0}$. To do so, using \eqref{eq:fried1zlate}, we write
\begin{equation}
\begin{aligned}
k_{\rm eq}=H_0\big[&\Omega_{\rm M,0}(1+z_{\rm eq})^{-2}+2\Omega_{\rm m,0}(1+z_{\rm eq})^{1+\frac{1}{1+\omega}}\\
&+\Omega_{\rm \sigma^2,0}(1+z_{\rm eq})^{4+\frac{2}{1+\omega}}\big]^{1/2},
\end{aligned}
\end{equation}
where we use $a_0=a(z=0)=1$, and $\Omega_{\rm M,0}=1-\Omega_{\rm m,0}-\Omega_{\rm r,0}-\Omega_{\rm \sigma^2,0}$. We note that, for $\omega\gtrsim 50$ and $z_{\rm eq }\sim 3369$ (see Table~\ref{tab:priors}), the contribution to $k_{\rm eq}$ from the term with $\Omega_{\rm M,0}$ is negligible, $k_{\rm eq}$ is by far more sensitive to $\Omega_{\rm \sigma^2,0}$ and, for a given set of values of the parameters, $k_{\rm eq}$ increases with increasing $\Omega_{\rm \sigma^2,0}$ and decreases with increasing $\omega$ (viz., as the BD gravity approaches to GR).

\begin{figure}[t]
\captionsetup{justification=raggedright,singlelinecheck=false,font=footnotesize}
\par
\begin{center}
   \includegraphics[trim = 1mm  1mm 1mm 1mm, clip, width=7.cm, height=4.5cm]{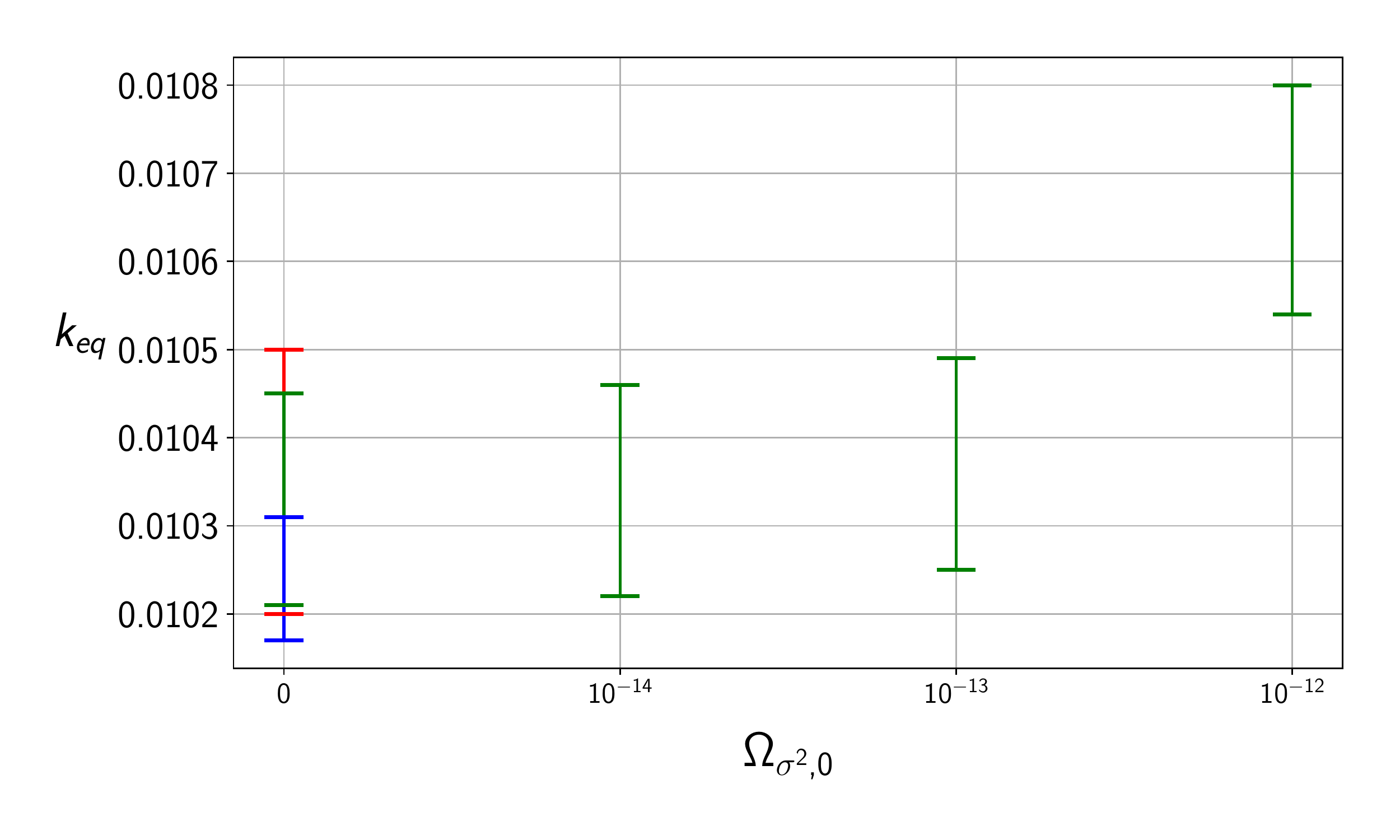}
\end{center}
\caption{The wavenumber ($k_{\rm eq}$) of a mode that enters the horizon ($H_{\rm eq}a_{\rm eq}$) at matter-radiation transition ($z_{\rm eq}$), viz., the maximum of matter power spectrum, for some fixed values of the expansion anisotropy today ($\Omega_{\sigma^2,0}$). Anisotropic BD extension of the $\Lambda$CDM model, is labeled by green color, the standard base $\Lambda$CDM (in this study) is labeled by blue color, while the recent Planck release value for the standard base $\Lambda$CDM model is shown with red color.}
\label{fig:keqani}
\end{figure}

We obtain $k_{\rm eq}=0.01034 \pm 0.00012$ for $\Omega_{\sigma^2,0}=0$ (isotropic expansion), which is slightly larger than, but yet consistent with the isotropic $\Lambda$CDM model value $k_{\rm eq}=0.01024 \pm 0.00007$. Accordingly, switching to the massive BD \eqref{eq:action} leads only to a slight increase in $k_{\rm eq}$, as expected from the constraint $\omega\gtrsim 50$ (see Table~\ref{tab:priors}). It may be useful to note that these two values are consistent with the recent Planck release \cite{Aghanim:2018eyx} value $k_{\rm eq}=0.010339\pm 0.000063$ (TT,TE,EE+lowE+lensing+BAO) obtained for the base $\Lambda$CDM model. On the other hand, we see that the expansion anisotropy, although negligible today, shifts $k_{\rm eq}$ to unrealistically large values, viz., we obtain $k_{\rm eq}= 0.15159 \pm 0.00327$ for the upper bound $\Omega_{\sigma^2,0}=10^{-8.48}$ (see Table~\ref{tab:priors}). We then work out the values of $\Omega_{\sigma^2,0}$ that can shift $k_{\rm eq}$ to reasonable values. See Fig.~\ref{fig:keqani} showing explicitly $k_{\rm eq}$ with respect to $\Omega_{\sigma^2,0}$ with errors to make a simple comparison between the models. We obtain $k_{{\rm eq}}=0.02824 \pm 0.00057$ by setting $\Omega_{\sigma^2,0}=10^{-10}$, and $k_{{\rm eq}}=0.01067 \pm 0.00013$ by setting $\Omega_{\sigma^2,0}=10^{-12}$, which are still inconsistent with the $k_{\rm eq}$ values given for the isotropic $\Lambda$CDM model. But then, we obtain $k_{{\rm eq}}=0.01037 \pm 0.00012$ by setting $\Omega_{\sigma^2,0}=10^{-13}$, and $k_{{\rm eq}}=0.01034 \pm 0.00012$ by setting $\Omega_{\sigma^2,0}=10^{-14}$, where we notice that only the last decimals are different. We observe further that $k_{\rm eq}$ does not change for $\Omega_{\sigma^2,0}\lesssim 10^{-14}$ anymore and remains consistent with the isotropic $\Lambda$CDM model values obtained in this study as well as recent Planck release, which in turn implies that we cannot distinguish $\Omega_{\sigma^2,0}\lesssim 10^{-14}$ from $\Omega_{\sigma^2,0}=0$ by means of $k_{\rm eq}$. We finally calculate using $\Omega_{\sigma^2,0}\sim 10^{-14}$ that the Universe is indeed dominated by matter+radiation at matter-radiation equality, viz., we now have $\frac{\Omega_{\sigma^2,\rm eq}}{2\Omega_{\rm m,eq}}\sim 10^{-3}$. Thus, by means of matter-radiation transition, we conclude
\begin{equation}
\Omega_{\sigma^2,0}\lesssim 10^{-14}\quad\textnormal{along with}\quad \omega\gtrsim 50,
\end{equation}
where the upper bound on the expansion anisotropy is improved by reducing about six orders of magnitude w.r.t. the ones given in Table~\ref{tab:priors}.

\subsection{Big Bang Nucleosynthesis}
\label{BBN} 
Big Bang Nucleosynthesis provides a probe of the dynamics of the early Universe, which in turn gives us opportunity to further investigate the constraints on the anisotropic BD extension of the standard $\Lambda$CDM model. Such that, in the standard-BBN (SBBN) --assuming the standard model of particle physics is valid and the expansion of the Universe is governed by GR-- the processes relevant to BBN take place when the temperature ranges from $T\sim 1\, {\rm MeV}$ to $T\sim 0.1\, {\rm MeV}$ and the age of the Universe from $t\sim 1\,{\rm s}$ to $t\sim 3\,{\rm min}$ corresponding to redshift $z\sim 3\times 10^8$ at which the Universe is radiation dominated. These, of course, should not be altered significantly in a viable cosmological model. Accordingly, we first looked for the condition of radiation dominance at $z\sim 3\times 10^8$ (implying $z_{\rm eq,\sigma^2,r}\gtrsim 3\times 10^8$) by using the constraints on the relevant parameters from Table~\ref{tab:priors} and found out that $\Omega_{\sigma^2,0}\lesssim 10^{-21}$ for $\omega\gtrsim 50$ in line with our preliminary investigation in Sect.~\ref{radsol} [cf., see Eq. \eqref{omsigma0constraint}]. On the other hand, $z_{\rm eq,\sigma^2,r}$ corresponds to the redshift at which $\rho_{\sigma^2}/\rho_{\rm r}=1$, but as may be seen from the investigations in \cite{Barrow:1976rda,Campanelli:2011aa} the expansion anisotropy does not lead to a considerable deviation from the SBBN for the  $\rho_{\sigma^2}/\rho_{\rm r}$ ratio up to a few percents, viz.,
\begin{equation}
\label{eqn:bbnstrongconst}
\frac{\rho_{\sigma^2}(z=z_{\rm BBN})}{\rho_{\rm r}(z=z_{\rm BBN})}\lesssim 10^{-2}.
\end{equation}
 Hence, considering this condition, we can obtain a new constraint on $\Omega_{\sigma^2,0}$ stronger than the one given in \eqref{omsigma0constraint} obtained from the condition $z_{\rm eq,\sigma^2,r}\gtrsim 3\times 10^8$. To work this out, we first write, from \eqref{eq:fried1zearly},
\begin{equation}
\begin{aligned}
\label{sigmaoverrhoagain}
\frac{\rho_{\sigma^2}}{\rho_{\rm r}} &\sim\frac{\rho_{\sigma^2,0}(1+z_{\rm eq})^{\frac{1}{1+\omega}}(1+z)^6}{\rho_{\rm r,0}(1+z)^4}\\
&\quad\quad\quad\quad\quad=\frac{\Omega_{\sigma^2,0}}{\Omega_{\rm m,0}}(1+z_{\rm eq})^{\frac{1}{1+\omega}+1}(1+z)^2,
\end{aligned}
\end{equation}
where we also used $\frac{\Omega_{\rm m,0}}{\Omega_{\rm r,0}}=\frac{\rho_{\rm m,0}}{\rho_{\rm r,0}}=1+z_{\rm eq}$. Then, we use in this equation the condition \eqref{eqn:bbnstrongconst} along with the constraints on $\Omega_{\rm m,0}$ and $z_{\rm eq}$ given in Table~\ref{tab:priors} by setting $z=3\times 10^8$ and reach the following constraint
\begin{equation}
\label{stronganisocond}
\Omega_{\sigma^2,0}\lesssim 10^{-23} \quad\textnormal{along with}\quad \omega\gtrsim 50.
\end{equation}
We note that, here, the contribution from the effective anisotropic pressure of the Jordan field during $z<z_{\rm eq}$, which is encoded in the term $(1+z_{\rm eq})^{\frac{1}{1+\omega}}$ in \eqref{sigmaoverrhoagain}, is not significant, such that it leads to only fifteen percent smaller upper bound value for $\Omega_{\sigma^2,0}$, viz., $(1+z_{\rm eq})^{-\frac{1}{1+\omega}}\lesssim 10^{-0.07}=0.85$ for $\omega\gtrsim 50$. Thus, provided that the condition on the expansion anisotropy today given in \eqref{stronganisocond} is satisfied, the expansion anisotropy will not have considerable effect on BBN, but the altered strength of the cosmological gravitation coupling depending on the BD parameter $\omega$ will, as we shall investigate in what follows.

 \begin{figure}[t!]
\captionsetup{justification=raggedright,singlelinecheck=false,font=footnotesize}
\par
\begin{center}
  \includegraphics[trim = 1mm  1mm 1mm 1mm, clip, width=6.5cm, height=4.5cm]{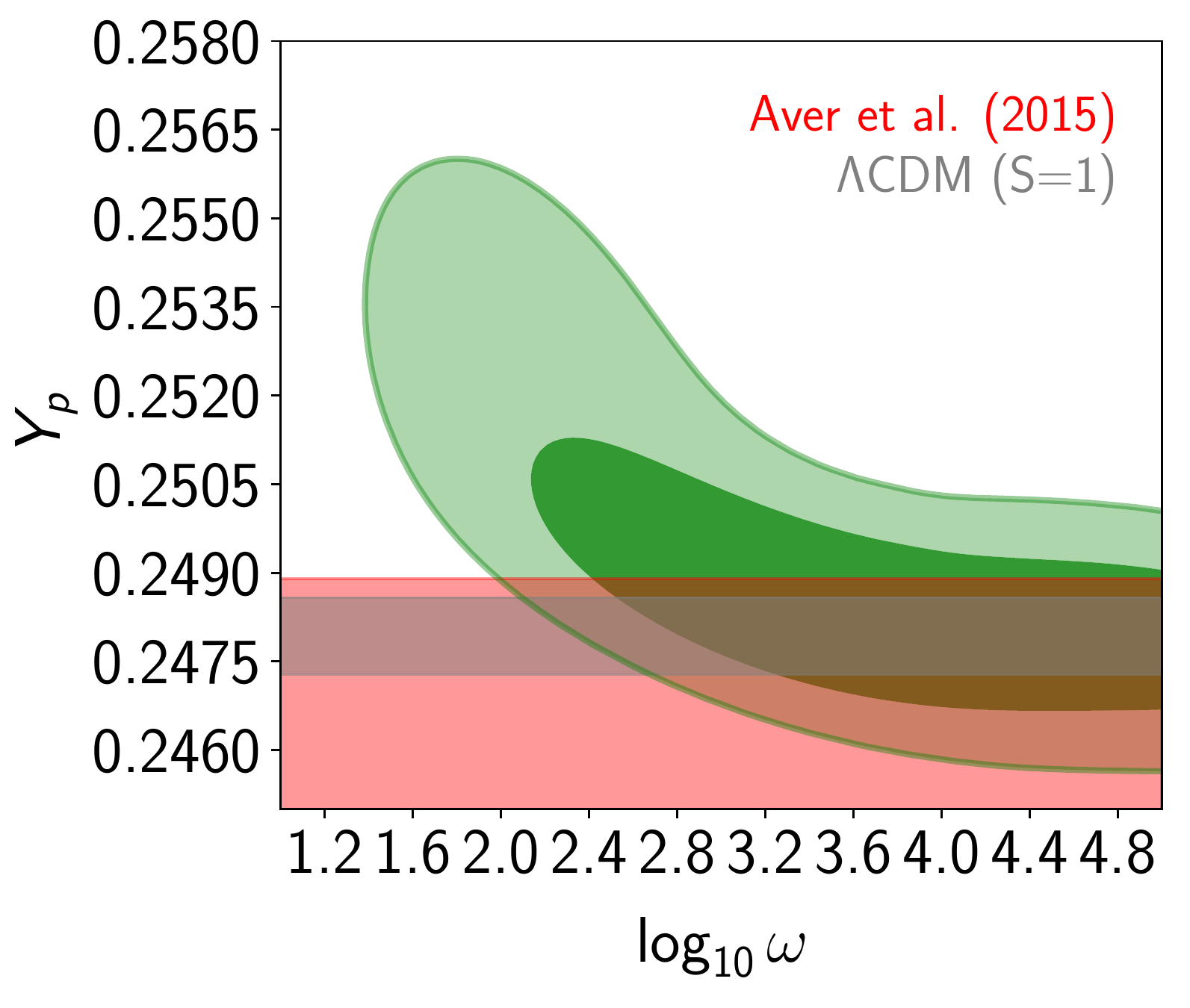}
   \includegraphics[trim = 1mm  1mm 1mm 1mm, clip, width=6.5cm, height=4.5cm]{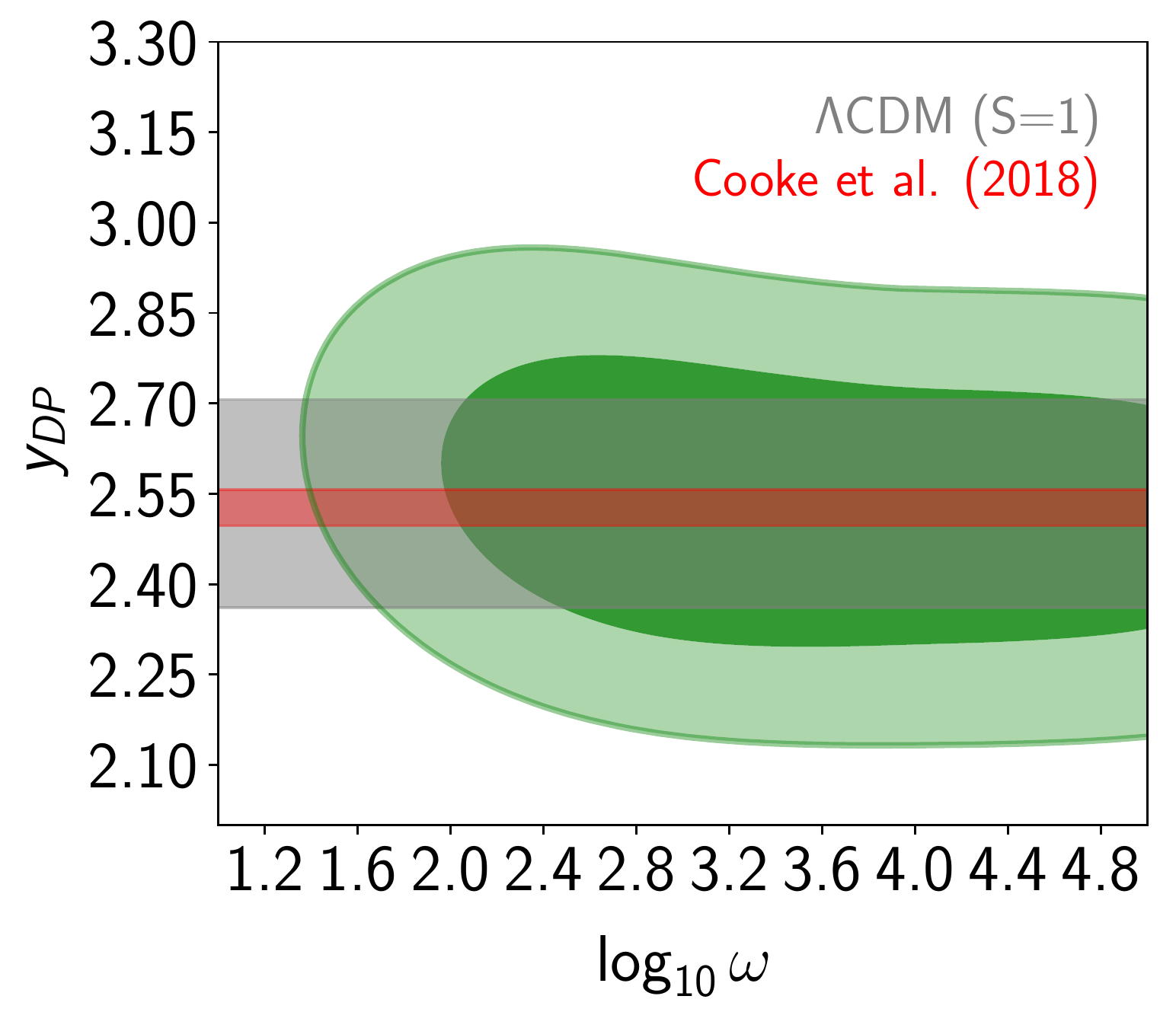}
\end{center}
\caption{The 2D constraints on the BBN predicted $^4$He (helium) mass fraction $Y_{\rm p}$ (top panel) and D (deuterium) $y_{\rm DP}$ mass fraction (bottom panel) in BD gravity are plotted with 1$\sigma$ (dark green) and 2$\sigma$ (light green) confidence contours. Grey bands are for the SBBN-predicted mass fractions with 1$\sigma$. Red bands are independent observational estimates  with 1$\sigma$  from \cite{Aver:2015iza,Cooke:2017cwo}. }
\label{fig:Ys}
\end{figure}

Provided that the condition \eqref{stronganisocond} is satisfied, the Universe is radiation dominated during the BBN epoch and hence, as it is discussed in Sect.~\ref{radsol}, we have $a\propto t^{\frac{1}{2}}$ as in the SBBN based on GR except that the strength of the cosmological gravitational coupling during this epoch, $G_1$, can be slightly larger than its present time value, $G_0$, in our model based on BD gravity, see \eqref{G1G0relnum} and \eqref{G1G0rel}. Consequently, in accordance with \eqref{G1G0rel}, we can write the expansion rate of the Universe during the radiation dominance, hence during the BBN as well, as follows:
\begin{equation}
\label{modHG}
H^2=\frac{G_1 }{G_0}H^2_{\rm SBBN}\quad\textnormal{with}\quad H^2_{\rm SBBN}=\frac{8\pi G_0 }{3}\frac{\pi^2}{30}g_{*}T^4,
\end{equation}
where $T$ is the temperature and $g_{*}(T)$ is the effective number of degrees of freedom counting the number of relativistic particle species determining the energy density in radiation as $\rho_{\rm r}=\frac{\pi^2}{30}g_{*}T^4$. According to this, BD gravity can lead to a larger expansion rate for a given temperature since $G_1>G_0$ is allowed, see \eqref{G1G0relnum}. We note that this is analogue of altering the expansion rate of the Universe during BBN by modifying $g_*$ (for instance, by introducing an extra massless degree of freedom such as sterile neutrino) within GR as
\begin{equation}
\label{modH}
H^2=\frac{\tilde g_* }{g_*}H^2_{\rm SBBN} \quad\textnormal{with}\quad H^2_{\rm SBBN}=\frac{8\pi G_0 }{3}\frac{\pi^2}{30}g_{*}T^4.
\end{equation}
It is clear from \eqref{modHG} and \eqref{modH} that we can set the following relation $\tilde g_*=\frac{G_1}{G_0} g_{*}$ implying that the altered $G$, i.e., $G_1$, in the BD gravity can equivalently be treated as modified $g_*$, namely, $\tilde g_*$, in GR. The effect of altered expansion rate of the Universe during BBN due to the modified $g_*$ within GR is well investigated in the literature \cite{Kneller:2004jz,Steigman:2007xt,Steigman:2012ve} and is parametrized in terms of $S\equiv\frac{H}{H_{\rm SBBN}}=\sqrt{\frac{\tilde g_*}{g_*}}$, which can be adopted as $S=\sqrt{\frac{G_1}{G_0}}$ for our work within BD gravity by preserving $g_*$ as in the SBBN. Such that, deviations from $S=1$ (SBBN) will modify the neutron abundance and the time available for nuclear production/destruction, changing the BBN-predicted primordial element abundances. In general, it is necessary to access to a BBN code to study the BBN-predicted primordial abundances (viz., mass fractions) as functions of $S$ and $\eta_{10}$ (the number density of baryons $n_{\rm b}$ to the number density of CMB photons $n_{\rm \gamma}$ defined as $\eta_{10}=10^{10} n_{\rm b}/n_{\rm \gamma}$ ). On the other hand, luckily, they have been identified (e.g., in  \cite{Kneller:2004jz,Steigman:2007xt,Steigman:2012ve}) by extremely simple but quite accurate analytic fits over a limited range in these variables as $5.5\lesssim \eta_{10}\lesssim 6.5$ and $0.85\lesssim S \lesssim 1.15$ and, for $^4$He (helium) and D (deuterium) these are\footnote{We consider $Y_{\rm p}$ and $y_{\rm DP}$ equations given in Ref. \cite{Steigman:2012ve}, which were updated with respect to the ones given in \cite{Kneller:2004jz,Steigman:2007xt} in accordance with, e.g, most importantly, the change in the recommended neutron lifetime from $\tau_{\rm n}=885.7\pm0.8\,{\rm s}$ to $\tau_{\rm n}=881.5\pm1.5\,{\rm s}$ by Particle Data Group in 2011 \cite{Nakamura:2010zzi}. To adopt these equations from \cite{Steigman:2012ve} for our work, as we assume the standard particle physics is valid, we set $\xi=0$ in the original equations given in \cite{Steigman:2012ve}, where $\xi$ is a parameter to quantify the lepton asymmetry in non-standard particle physics.} \cite{Steigman:2012ve}
\begin{align}
\label{eqnyp}
Y_{\rm p}=0.2381\pm0.0006+0.0016[\eta_{10}+100 (S-1)],\\
y_{\rm DP}=45.7(1\pm0.06)[\eta_{10}-6(S-1)]^{-1.6},
\label{eqnydp}
\end{align}
where, $Y_{\rm p}\equiv\frac{n_{\rm He}}{n_{\rm b}}$ and $y_{\rm DP}\equiv 10^5\frac{n_{\rm D}}{n_{\rm  H}}$ are $^4$He and D mass fractions, respectively, and here  in our study, $\eta_{10}=273.9\,\Omega_{\rm b,0} h^2$ and $ S=\sqrt{\frac{G_1}{G_0}}\simeq(1+z_{\rm eq})^{\frac{1}{2(1+\omega)}}$, see \eqref{G1G0rel}. We see that $\eta_{10}\approx6.1$ for both models from the constraints on $\Omega_{\rm b,0} h^2$ given in Table~\ref{tab:priors} and find that the cosmological data we considered constrain the altered expansion rate of the Universe for $z>z_{\rm eq}$, hence during BBN, as $1<S<1.0665$ at $68\% {\rm \, C.L.}$  and $1<S<1.0839$ at $95\% {\rm \, C.L.}$. These are well inside the validity intervals of \eqref{eqnyp} and \eqref{eqnydp} and hence they can be safely utilized here for obtaining the BBN-predicted $Y_{\rm p}$ and $y_{\rm DP}$ values, which in turn can be used for a further investigation of the constraint on the BD parameter $\omega$.

In the standard $\Lambda$CDM model [$S=1$ (GR), $\Omega_{\sigma^2,0}=0$ and accommodating SBBN],
considering the constraint on $\Omega_{\rm b,0}h^2$ given in Table~\ref{tab:priors}, we obtain $\eta_{10}=6.1469 \pm 0.0404$ leading, from \eqref{eqnyp} and \eqref{eqnydp}, to
\begin{align}
\label{eqn:BBNpredictGR}
Y_{\rm P}^{\rm SBBN}&=0.24793\pm 0.00065\quad (68\%\, {\rm C.L.}),\\
y_{\rm DP}^{\rm SBBN}&=2.4999 \pm 0.1642\quad (68\%\, {\rm C.L.}).
\end{align}
On the other hand, in the BD extension of the standard $\Lambda$CDM model, by assuming $\Omega_{\sigma^2,0}\lesssim 10^{-23}$ [viz., expansion anisotropy is negligible during BBN, see \eqref{stronganisocond}] and considering the constraints on $\Omega_{\rm b,0}h^2$, $\omega$ and $z_{\rm eq}$ given in Table~\ref{tab:priors}, we obtain $\eta_{10}=6.1373\pm 0.0448$ and $1<S<1.0665$ at $68\% {\rm \, C.L.}$ leading, from \eqref{eqnyp} and \eqref{eqnydp}, to
\begin{align}
\label{eqn:BBNpredictBD}
Y_{\rm p}&=0.24898\pm 0.00199\quad (68\%\, {\rm C.L.}),\\
y_{\rm DP}&=2.5338\pm 0.1726\quad (68\%\, {\rm C.L.}).
\end{align}
 Observations of helium and hydrogen recombination lines from metal-poor extragalactic H II regions provide an independent method (viz., a direct measurement) for determining the primordial helium abundance and a latest and widely accepted estimate comes from the data compilations of \cite{Aver:2015iza} giving $Y_{\rm p}=0.2449 \pm 0.0040$ (68 $\%$ C.L.). Similarly, the most recent estimate of the primordial deuterium abundance comes from the best seven measurements in metal-poor damped Lyman-$\alpha$ systems studied in \cite{Cooke:2017cwo} giving $y_{\rm DP}=2.527\pm 0.030$ (68 $\%$ C.L.). We note that the BBN-predicted $Y_{\rm p}$ and $y_{\rm DP}$ for both the $\Lambda$CDM model \eqref{eqn:BBNpredictGR} and its anisotropic BD extension \eqref{eqn:BBNpredictBD} obtained by using the cosmological data, led to consistent values with independent observational estimates. We notice, however, slightly larger mean values in the case of BD gravity, as suggested by \eqref{eqnyp} and \eqref{eqnydp} when $S>1$ (assuming $\eta_{10}$ is fixed). We give a summary of our findings by depicting the 2D marginalized posterior distribution of the BBN-predicted $Y_{\rm P}$ via \eqref{eqnyp}/$y_{\rm DP}$ via \eqref{eqnydp} and $\omega$ in the top panel/bottom panel of Fig.~\ref{fig:Ys}. In which, for a comparison, we depict also the bands of the SBBN-predicted $Y_{\rm P}$ via \eqref{eqnyp}/$y_{\rm DP}$ via \eqref{eqnydp} for the $\Lambda$CDM model accommodating SBBN ($S=1$) and of their independent observational estimates given in \cite{Aver:2015iza,Cooke:2017cwo}. We note that there is an increasing anti-correlation between $Y_{\rm P}$ and $\omega$ with decreasing $\omega$ as it approaches its lower bound, whereas for large values of $\omega$ the $Y_{\rm P}-\omega$ contour at 68\% C.L. (dark green) approaches the band of the SBNN-predicted $Y_{\rm P}$ at 68\% C.L. (grey) as it should be. Besides these, more importantly, we see that the $Y_{\rm P}-\omega$ contour at 68\% C.L. (dark green) stays above the band of the independent observational estimate band (red) for $\omega\lesssim 250$, namely, its consistency with the independent observational estimate of $Y_{\rm p}=0.2449 \pm 0.0040$ from \cite{Aver:2015iza} requires $\omega\gtrsim 250$ at 68\% C.L. providing us an improved constraint on the BD parameter $\omega$. We see that the BBN-predicted $y_{\rm DP}$ via \eqref{eqnydp} is insensitive to  $\omega$, $y_{\rm DP}-\omega$ contour at 68\% C.L. (dark green) already covers the independent observational estimate band (red) for $\omega\gtrsim250$ (the improved constraint from our $Y_{\rm P}$ investigation) but is much wider than it (due to the relatively large internal error in \eqref{eqnydp}, viz., $y_{\rm DP}\propto 1\pm0.06$), and $y_{\rm DP}-\omega$ contour for BD gravity (dark green) approaches the SBBN-predicted $y_{\rm DP}$ band at 68\% C.L. (grey) for large $\omega$ values as it should be. These show that we are not able to deduce a new constraint on the BD parameter $\omega$ by comparing the BBN-predicted value of $y_{\rm DP}$ via \eqref{eqnydp} with its independent observational estimate.
 
 Thus, by means of the BBN, we reach the most stringent constraints on $\Omega_{\sigma^2,0}$ and $\omega$, the two free parameters that determine the anisotropic BD extension to the standard $\Lambda$CDM model, as follows;
\begin{equation}
\Omega_{\sigma^2,0}\lesssim 10^{-23}\quad\textnormal{and}\quad\omega\gtrsim 250.
\end{equation}
Finally, considering these improved constraints, we find that the contribution to this constraint of $\Omega_{\sigma^2,0}$ from the anisotropy of the effective pressure of the Jordan field during $z<z_{\rm eq}$, which is encoded in the term $(1+z_{\rm eq})^{\frac{1}{1+\omega}}$ in \eqref{sigmaoverrhoagain}, is quite insignificant, such that, it leads to only a few percent smaller upper bound value for $\Omega_{\sigma^2,0}$, viz., $(1+z_{\rm eq})^{-\frac{1}{1+\omega}}\lesssim 10^{-0.01}=0.977$ for $\omega\gtrsim 250$.

\section{Conclusions}
\label{sec:closing}

We have carried out an explicit detailed theoretical and observational investigation of an anisotropic massive Brans-Dicke (BD) gravity \textit{extension} of the standard $\Lambda$CDM model, wherein the \textit{extension} is characterized by the two additional degrees of freedom: the so called BD parameter $\omega$ and the present day value of the density parameter corresponding to the shear scalar (a measure of the expansion anisotropy), $\Omega_{\sigma^2,0}$. The role of the cosmological constant $\Lambda$ is taken over by the Jordan field potential of the form $U(\varphi)=\frac{1}{2}M^2\varphi^2$ with $M$ being the bare mass of the field. We have considered the LRS Bianchi type I metric, which generalizes the spatially flat RW metric simply by allowing a different scale factor along one of the three orthogonal axes, while preserving the spatial homogeneity and flatness, and the isotropic spatial curvature\footnote{In more complicated anisotropic spacetimes, the anisotropic spatial curvature contributes to the shear propagation equation and hence to the redshift dependency of the shear scalar. For instance, the most general spatially flat anisotropic spacetimes are of Bianchi type VII$_0$, which, in addition to the simple expansion-rate anisotropies, yield anisotropic spatial curvature that mimics traceless anisotropic fluid \cite{Barrow:1997sy}.}. We have solved the field equations analytically and obtained the average Hubble parameter, $H(z)$, explicitly by extending the method developed in \cite{Boisseau:2010pd}; described the anisotropic effective DE in accordance with the way of defining effective source given in \cite{Boisseau:2000pr,Gannouji:2006jm}, and consistently included the radiation into the model.

The BD parameter $\omega$, being the measure of the deviation from GR ($|\omega|\rightarrow\infty$), by alone characterizes the dynamical behaviour of the effective DE as well as the redshift dependency of the expansion anisotropy. These two  affect each other depending on $\omega$, such that the shear scalar contributes to the dynamics of the effective DE, and its anisotropic stress controls the dynamics of the shear scalar, in particular deviations from its usual form $\sigma^2\propto(1+z)^6$ in GR, see $H(z)$ in \eqref{eq:fried1zlate}. We planned the current study as an \textit{extension} in the sense of a \textit{correction} to the standard $\Lambda$CDM model, so we have mainly confined our investigations to small deviations from this model via sufficiently small $\Omega_{\sigma^2,0}$ and large $\omega$ values. We have shown through some  preliminary cosmological discussions that $|\omega|\lesssim10$ (roughly) cannot be called as a small deviation, yet, for completeness we have extended our investigations to non-negative values of $\omega$. Indeed, considering the combined data sets PLK+BAO+SN+$H$, we have obtained $\omega\gtrsim50$, $M<1.51\times 10^{-34}\rm eV$ and $\Omega_{\sigma^2,0}\lesssim 10^{-8.48}$ ($\Omega_{\sigma^2,0}\lesssim 10^{-14}$, when matter-radiation equality is considered).\footnote{We make use of a modified version of the simple and fast MCMC code that computes expansion rates and distances from the Friedmann equation, named SimpleMC \cite{Anze, Aubourg:2014yra}. The method we use constrains the expansion anisotropy through its contribution to the average expansion rate of the Universe.} Then by means of BBN, we have improved these constraints to $\omega\gtrsim250$ (in particular, from the comparison of the helium abundance prediction of the model with the direct measurements) and $\Omega_{\sigma^2,0}\lesssim 10^{-23}$.  We have also found that the contribution of the anisotropy of the effective DE (viz., $\Delta w_{\rm DE,0}<4.23\times 10^{-7}$) to this constraint on $\Omega_{\sigma^2,0}$ is insignificant, namely, led to only a few percent smaller (stronger) upper bound value.\footnote{It can be seen from $H(z)$ given in \eqref{eq:fried1zlate} that the sufficiently large negative $\omega$ values also lead to small deviations from $\Lambda$CDM, but in this case we have slightly flatter redshift dependence of the expansion anisotropy, which in turn can relax the constraint on $\Omega_{\sigma^2,0}$ at few percents level compared to the one we have obtained along with large positive $\omega$ values. Nevertheless, for such large negative values of $\omega$, being less than the scale invariant limit $-\frac{3}{2}$, one also should cope with stability issues.}

All these have led us to conclude that, with the observations relevant to the dynamics of the Universe, the simplest anisotropic massive BD gravity extension of the standard $\Lambda$CDM model presents no significant deviations from it all the way to the BBN. The strongest cosmological constraints on $\omega$ present in the literature, e.g., $\omega\gtrsim 890$ \cite{Avilez:2013dxa}, would obviously just strengthen this conclusion. Moreover, we should further consider the strongest local constraint $\omega>40000$ \cite{Bertotti:2003rm} as well since the constraint $M<1.51\times 10^{-34}\,\rm eV$ obtained here for the mass of the Jordan field leads to no relaxation on it \cite{Perivolaropoulos:2009ak}. This in turn implies that, against the local constraints, the cosmological features that arise from replacing GR by massive BD (such as the modified expansion anisotropy due to the anisotropic effective DE) are not significantly sensitive to the current cosmological observations. In other words, in view of  the local constraints on $\omega$, the simplest anisotropic massive BD gravity extension of the standard $\Lambda$CDM model cannot be distinguished from its simplest GR based anisotropic extension \cite{Akarsu:2019pwn} when we consider the data from cosmological observations.\footnote{For example, the deviation of the effective EoS parameter corresponding to the expansion anisotropy from $w_{\sigma^2}=1$ (corresponding to the GR limit) becomes just $\sim10^{-5}$ for $\omega=40000$, whilst it is already very small, viz., $\sim10^{-3}$, for the cosmological lower bound $\omega\sim250$ (and $\omega\sim890$).}

It is conceivable that our findings, when we consider only the observations relevant to the background dynamics of the Universe, are representative for the ones that would be obtained in the more general extensions, namely, through a gravity theory more general than the massive BD and therefore can evade the local constraints. For instance, promoting $\omega$ to be some functions of $\varphi$, the observations from the Cassini spacecraft place upon strong constraint on $\omega$ as in the original BD theory. This constraint, however, now only applies to the local value of $\omega(\varphi)$, i.e., of the present day value of $\varphi$ in the Solar System. Similarly, it is possible that $\omega$ has spatial variation, so that, for instance, it could be constrained to be $\omega>40000$ locally, but could be in line with $\omega\sim 10^2$ at cosmological scales \cite{Barrow:1999qk,Clifton:2004st}. See, for instance, \cite{Clifton:2011jh} and references therein for a further reading on such gravity theories closely related to the BD theory, of which the Horndeski/Galileon \cite{Horndeski:1974wa,Deffayet:2011gz,Nicolis:2008in} theories are the most popular ones. In the anisotropic extensions through such gravity theories, however, one should deal with more complicated field equations, those may not be solved analytically and lead to the explicit expression of $H(z)$, and hence the solutions and findings, presented in this work, are useful as they may be considered as the good approximations to such constructions. Consequently, the theoretical investigations we carried out here are, in general, instructive for the anisotropic extensions of the standard $\Lambda$CDM model replacing GR by a modified theory of gravity approximating massive BD at cosmological scales. When the observational constraints are considered as well, our findings against the cosmological constraints (the local constraints) with $\omega\gtrsim 10^{2}$ ($\omega\gtrsim 40000$) are instructive for the extensions through those gravity theories that can (cannot) evade local constraints. Yet, this also implies that the lesson learned from this study based on the massive BD would approximately be valid for such more general constructions, namely, many interesting features of such anisotropic extensions of the $\Lambda$CDM model that do not exist within its simple GR based anisotropic extension, would remain insignificant when the model is constrained by the observations.

We note that, in fact, these features  become quite significant for $|\omega|\sim O(1)$, which, interestingly, is in principle the natural order of magnitude for the BD parameter (as, e.g., $\omega=-1$ appears in the low energy limit of string theories). This by alone, makes it quite interesting, at least theoretically, to further study this region in spite of that, as we have discussed above, it implies large deviations from the standard $\Lambda$CDM dynamics, and hence is not expected to be consistent with the real Universe. Such a study, however, is beyond the scope of the current paper, but yet, here, we would like to make a couple of relevant comments. In case of small negative values of $\omega$, say, $\omega\sim-1$ but larger than the scale invariant limit $-\frac{3}{2}$ (below which stability issues appear) the model exhibits several critical points that complicates the investigation in this region. So it is necessary to carry out several separate analyses within this small region. Meanwhile, this region is distinguished that it is relevant to string theories, namely, $\omega=-1$ corresponds to the low energy effective string action and $\omega\sim-1$ appears in $d$-brane constructions \cite{Duff:1994an,Lidsey:1999mc,Callan:1985ia,Fradkin:1985ys}. For these reasons in fact, although we have devoted the current work to non-negative $\omega$ values, we carried out a discussion in Sect.\ref{anidiscuss} showing how dramatically the redshift dependence of the expansion anisotropy can alter when $\omega \sim-1$. For instance, for $\omega=-\frac{4}{3}$, expansion anisotropy becomes non-dynamical (mimicking $\Lambda$) and for $\omega=-1$, expansion anisotropy should be set to zero. The region $-\frac{3}{2}<\omega<-\frac{4}{3}$ is also interesting that, in the presence of only dust (without $\Lambda$ or the mass of Jordan field), it is the region of BD theory leading to accelerated expansion of the Universe, and we notice that the expansion anisotropy also contributes to this acceleration since it also behaves like DE in this region. We do not know, so far, a concrete and successful mechanism making such small values (negative or positive) of $\omega$ consistent with local experiments and cosmological observations. Yet, involving extra spatial dimensions may be good place for looking for such possibilities as it was shown that it is possible to make BD theory (massive or massless) consistent with the gravitational tests (including solar system tests) for $|\omega|=O(1)$ in the presence of extra dimensions \cite{Akarsu:2019oem}. Noticing the presence of extra dimensions as part of string theories and the dynamical extra dimensions (internal space) contribute to the evolution of the four dimensional anisotropic spacetime like an anisotropic stress in a similar way that the Jordan field does in BD theory make this option more appealing and interesting \cite{Rasouli:2011rv,Rasouli:2014dxa,Rasouli:2019wzw,Maartens:2000az,Campos:2003bj}. This shows one of the many possibilities of rich behaviors which could be worthwhile to investigate in future studies on the extensions of the standard $\Lambda$CDM model wherein the expansion anisotropy exhibits non-trivial behaviors that may have interesting cosmological consequences.

A more rigorous observational investigation of the model may be another direction to follow. We have studied the observational constraints on the parameters of the model by considering the background dynamics of the Universe through the evolution of the comoving volume scale factor. These constraints can be improved when we consider the perturbation sector --as in \cite{Avilez:2013dxa,Ballardini:2016cvy,Umilta:2015cta} providing the most stringent cosmological constraints on $\omega$-- along with, for instance, the full Planck data --as in \cite{Saadeh:2016bmp,Saadeh:2016sak} providing constraints on the anisotropic expansion on the top of the $\Lambda$CDM at a level close to the ones from BBN by considering the CMB radiation temperature and polarization data-. Such a further observational investigation of the model requires considerable amount of work beyond the aim of the current paper, as it is necessary to study perturbations on the anisotropic background in the presence of an effective anisotropic source. This would strengthen the constraints on the parameters of the model, but would probably not change our conclusion that the model exhibits no significant deviations from the standard $\Lambda$CDM model all the way to the BBN. Yet, we cannot be sure about that unless we undertake this work in future and see the results.

\begin{acknowledgements}
The authors thank to Shahin Sheikh-Jabbari, Mehmet \"{O}zkan and Suresh Kumar for valuable discussions. \"{O}.A. acknowledges the support of the Turkish Academy of Sciences in scheme of the Outstanding Young Scientist Award  (T\"{U}BA-GEB\.{I}P). \"{O}.A. is grateful for the hospitality of the Abdus Salam International Center for Theoretical Physics (ICTP) while the part of this research was being carried out. N.K. acknowledges the post-doctoral research support from the {\.I}stanbul Technical University (ITU). J.A.V. acknowledges the support provided by the grants FOSEC SEP-CONACYT Investigaci\'on B\'asica A1-S-21925, and UNAM-PAPIIT IA102219. 
\end{acknowledgements}

\bigskip

\end{document}